\documentclass[a4paper,11pt]{article}
\usepackage{amsmath}
\usepackage[T1]{fontenc}
\usepackage{amssymb}
\usepackage{amsthm}

\newtheorem{thm}{Theorem}[section]
\newtheorem{prop}[thm]{Proposition}
\newtheorem{lem}[thm]{Lemma}

\newtheorem{proposition}[thm]{Proposition}

\newtheorem{rem}[thm]{Remark}

\numberwithin{equation}{section}
\font\twlmsb=msbm10 at 12pt \font\egtmsb=msbm8\font\sixmsb=msbm6
\newfam\msbfam
\textfont\msbfam=\twlmsb \scriptfont\msbfam=\egtmsb
\scriptscriptfont\msbfam=\sixmsb 
\title{{\bf On the stability of normal states for a generalized
Ginzburg-Landau model}}

\author{Ayman Kachmar\\
{\it Universit\'e de Paris-Sud,}\\
{\it D\'epartement de math\'ematiques,}\\
{\it B\^at. 425, 91405 Orsay, France}\\
{\it E-mail~:}~{ayman.kachmar@math.u-psud.fr}}
\date{October 18, 2006}

\begin{document}
\maketitle
\begin{abstract}
We formulate a spectral problem related to the onset of
superconductivity for a generalized Ginzburg-Landau model, where the
order parameter and the magnetic potential are defined in the whole
space. This model is devoted to the `proximity effect'  for a
superconducting sample surrounded by a normal material.
In the regime when the Ginzburg-Landau parameter (of the
superconducting material) is large, we estimate the critical applied
magnetic field for which the normal state will lose its stability, a
result that has some roots in the physical literature. In some
asymptotic situations, we recover results related to the `standard'
Ginzburg-Landau model, where we mention in particular the two-term
expansion for the upper critical field obtained by Helffer-Pan.
\end{abstract}
\markright{A. Kachmar,\quad Stability of normal states for
generalized G-L}

\tableofcontents
\newpage
\section{Introduction and main results}
\subsection{The generalized Ginzburg-Landau model}
It is predicted by the physicist de\,Gennes \cite{deGe, deGe1} that
the presence of a normal material exterior to a superconductor will
push the superconducting electron Cooper pairs to flow through the
normal material and to penetrate significatively over a band of
length $\frac1{b}$, called the `extrapolation length'. To understand
this phenomenon, which is called by de\,Gennes the `proximity
effect', one has to consider a generalized Ginzburg-Landau theory
where the order parameter and the magnetic potential are both
defined in the whole
space.\\
In the Ginzburg-Landau theory (cf.~\cite{GL}), the superconducting
properties are described by a complex valued wave function $\psi$,
called the `order parameter', and a real vector field $\mathcal A$,
called the `magnetic potential'. The pair $(\psi, A)$ has the
following physical interpretation~: $|\psi|^2$ measures the density
of the electron Cooper pairs (in particular, $\psi\equiv0$
corresponds to a normal state) and  ${\rm curl}\, A$ measures the
induced magnetic field in the sample. For cylindrical
superconductors with infinite height and placed in an applied
magnetic field parallel to the axis of the cylinder, it is
sufficient to define the pair $(\psi,A)$ on $\mathbb R^2$ (i.e. on
the 2-D cross section). The system is in equilibrium when the pair
$(\psi,A)$ minimizes the `Gibbs free
energy'.\\
After a proper scaling, the `Gibbs free energy'  has the following
form (cf.~\cite{Chetal})~:
\begin{equation}\label{E-T}
(\psi,A)\mapsto\mathcal G(\psi,A)=\int_{\mathbb R^2}
\left\{\frac1{\widetilde m}|\nabla_{\kappa HA}\psi|^2+\widetilde
a\kappa ^2|\psi|^2+\widetilde \beta\frac{\kappa^2}2|\psi|^4+(\kappa
H)^2|{\rm curl}\,A-1|^2\right\}dx,
\end{equation}
where we use the notation,
\begin{equation}\label{M-L}
\nabla_{\kappa HA}\psi=(\nabla-i\kappa HA)\psi.
\end{equation}
The functional (\ref{E-T}) depends on many parameters~: $\kappa>0$
is a temperature independent parameter called the `Ginzburg-Landau
parameter' (it is a characteristic of the superconducting material),
$H>0$ is the intensity of the constant applied magnetic field,
$\widetilde m,\widetilde a,\widetilde \beta$ are functions defined
in $\mathbb R^2$ and are depending on the material, temperature,
etc. Typically, the function $\widetilde a$ depends on the
temperature $T$ in the following way~:
$$\widetilde a\approx (T-T_c),$$
where $T_c$ is the critical temperature. As in \cite{Chetal}, we
take the functions $\widetilde m,\widetilde a,\widetilde \beta$ in
the following form~:
\begin{equation}\label{II-m,a,b}
\widetilde m= \left\{\begin{array}{l} 1,\text{ in }\Omega\\
m,\text{ in }\mathbb R^2\setminus\Omega
\end{array}\right.\quad
\widetilde a=\left\{\begin{array}{l} -1,\text{ in }\Omega\\
a,\text{ in }\mathbb R^2\setminus\Omega
\end{array}\right.\quad
\widetilde \beta=\left\{\begin{array}{l} 1,\text{ in }\Omega\\
0,\text{ in }\mathbb R^2\setminus\Omega.
\end{array}\right.
\end{equation}
Here $\Omega\subset\mathbb R^2$ is assumed to be open, bounded and
simply connected, and $a,m$ are positive constants. Typically,
$\Omega$ corresponds to a superconducting material\footnote{We
emphasize here that the Ginzburg-Landau parameter $\kappa$ is
determined {\it only} by the material in $\Omega$.} (i.e. below its
critical temperature) surrounded by a
normal material (i.e. above its critical temperature).\\
If $(\psi,A)\in H^1_{\rm loc}(\mathbb R^2; \mathbb C)\times H^1_{\rm
  loc}(\mathbb R^2;\mathbb R^2)$ is a critical point of $\mathcal G$,  then the following condition holds
for any $(\phi, B) \in C_0^\infty(\mathbb R^2;\mathbb C)\times
C_0^\infty(\mathbb R^2;\mathbb R^2)$,
\begin{eqnarray*}
&&\hskip-0.5cm\int_{\mathbb R^2} \left(\frac1{\widetilde
m}\Re\left(\nabla_{\kappa H A}\psi\cdot\nabla_{\kappa
  HA}\overline{\phi}\right)-\frac{\kappa H}{\widetilde m}B\cdot\Im\left(\overline{\psi}\nabla_{\kappa
  HA}\psi\right)+(\kappa H)^2{\rm curl
  }A\cdot{\rm curl} B\right)dx\\
&&=(\kappa H)^2\Re\left(\int_\Omega
(1-|\psi|^2)\psi\overline{\phi}\,dx-a\int_{\Omega^c}\psi\overline{\phi}\,dx\right).
\end{eqnarray*}
Thus, $(\psi,A)$ is a weak-solution of the following system of
equations, which we call `generalized Ginzburg-Landau equations',
\begin{equation}\label{GL-gen}
\left\{\begin{array}{l}
-\nabla_{\kappa HA}^2\psi=(\kappa H)^2(1-|\psi|^2)\psi,\quad \text{in }\Omega,\\
\\
{\rm curl}^2A=\frac1{\kappa
H}\Im\left(\overline{\psi}\nabla_{\kappa HA}\psi\right),
\quad \text{ in }\Omega,\\
\\
-\frac1{m}\nabla_{\kappa HA}^2\psi+a(\kappa H)^2\psi=0,\quad\text{ in }{\overline \Omega}^c,\\
\\
{\rm curl}^2A= \frac1{m(\kappa
H)}\Im\left(\overline{\psi}\nabla_{\kappa H A}\psi\right),\quad
\text{ in }
{\overline \Omega}^c,\\
\\
\mathcal T_{\partial\Omega}^{\rm int}\left\{\nu\cdot\nabla_{\kappa
HA}\psi\right\}= \frac1{\widetilde m}\mathcal
T_{\partial\Omega}^{\rm ext}\left\{\nu\cdot\nabla_{\kappa
HA}\psi\right\},
\quad \mathcal T_{\partial\Omega}^{\rm int} \psi=\mathcal T_{\partial\Omega}^{\rm ext}\psi,\\
\\
\mathcal T_{\partial\Omega}^{\rm int}\left({\rm curl
}\,A\right)=\mathcal T_{\partial\Omega}^{\rm ext} \left({\rm curl
A}\,\right),\quad \mathcal T_{\partial\Omega}^{\rm
int}\left(A\right)=\mathcal T_{\partial\Omega}^{\rm ext}\left(
A\right)\quad \text{ on }\partial\Omega.
\end{array}\right.
\end{equation}
In the above equations, $\nu$ is the unit outward normal vector of
$\partial\Omega$. We use the notations  $\mathcal T_{\partial\Omega}^{\rm int}$ and
$\mathcal T_{\partial \Omega}^{\rm ext}$ to denote, respectively, the `interior' and the `exterior' trace on $\partial\Omega$~:
$$\mathcal T_{\partial \Omega}^{\rm int}~: H^1(\Omega)\longrightarrow L^2(\partial\Omega),\quad
\mathcal T_{\partial \Omega}^{\rm ext}~:
H^1(\Omega^c)\longrightarrow L^2(\partial\Omega).$$
Although an increasing number of mathematicians become interested in
the problems arising from superconductivity, very few attention is
paid to  the functional (\ref{E-T}). In the former literature, the
authors are either concerned with the minimization of an  energy
functional defined only in $\Omega$ (cf. (\ref{E-s})), or they
replace the energy of the normal material  by a boundary term (cf.
(\ref{k2Re1-eq2})). We mention here for instance the works of
Bernoff-Sternberg~\cite{BeSt}, Baumann-Phillips-Tang~\cite{BaPhTa},
Lu-Pan~\cite{LuPa1}, Helffer-Morame~\cite{HeMo3},
Helffer-Pan~\cite{HePa}, Fournais-Helffer~\cite{FoHe} and
Kachmar~\cite{Kach}. Other `generalized' energy functionals similar
to (\ref{E-T}) and that models `inhomogeneous' superconducting
samples were also analyzed previously, see for instance
\cite{AlBron, AnBaPh} and the references therein. The inhomogeneity
there is only due to variations of the `critical temperature' within
the sample (i.e. $\widetilde m=1$ and $\widetilde a$ varies), and
the authors were mainly concerned with the analysis of `pinning
effects', that is, roughly speaking, the attraction of vortices
towards the less superconducting regions (unlike our situation where
we are concerned with the analysis of the onset of  superconductivity).\\
Perhaps it is in~\cite{Chetal} that the functional (\ref{E-T}) is
first introduced in the mathematical literature, but the analysis
there seems to remain  at a formal level. Recently, a rigorous
analysis of the functional (\ref{E-T}) has been carried out by
Giorgi~\cite{Gi}. Among other things, the author proves the
existence of minimizers for (\ref{E-T}) in a suitable functional
space (related to the analysis of Laplace's equation in $\mathbb
R^2$, see \cite{Ametal}), the existence of normal states and of an
upper critical field. By a normal state we mean a solution of
(\ref{GL-gen}) of the form $(0,\mathbf F)$. In our situation, we can
choose $\mathbf F$ in the following canonical way~:
$$\mathbf F(x_1,x_2)=\frac12(-x_2,x_1),\quad\forall (x_1,x_2)\in\mathbb R^2,$$
and we notice that $\mathbf F$ satisfies~:
$${\rm curl}\,\mathbf F=1,\quad {\rm div }\,\mathbf F=0,\quad\text{in }\mathbb R^2.$$ With the
above property, the vector field $\mathbf F$ is unique up to a gauge
transformation (cf. \cite{Gi}).

\subsection{Statement of the results}
A normal state is said to be stable\footnote{Our definition of
stability is actually that of `local stability', but in the regime
considered in this paper, we expect that `stability' and `local
stability' of normal states will coincide.} if it is a local minimum
of (\ref{E-T}). The Hessian of (\ref{E-T}) at the normal state
$(0,\mathbf F)$ is given by~:
$$\mathbb E\ni(\phi,B)\mapsto 2\mathcal Q[\kappa,H](\phi)+2(\kappa H)^2
\int_{\mathbb R^2}|{\rm curl}\,B|^2dx,$$ where the quadratic form
$\mathcal Q[\kappa,H]$ is defined by~:
\begin{equation}\label{qf}
\mathcal Q[\kappa,H](\phi)=\int_{\Omega}\left(|\nabla_{\kappa
H\mathbf F}\phi|^2-\kappa^2|\phi|^2\right)dx+
\int_{\Omega^c}\left(\frac1{m}|\nabla_{\kappa H\mathbf
F}\phi|^2+a\kappa^2|\phi|^2\right)dx.
\end{equation}
Thus, for $(0,\mathbf F)$ to be stable, a necessary condition
is to have~:
$$\mathcal Q[\kappa,H](\phi)\geq0,\quad\forall \phi\in\mathcal
H_{\kappa H\mathbf F}^1(\mathbb R^2),$$ where, given a vector field
$\mathcal A$ and an open set $U\subset\mathbb R^2$, the space
$\mathcal H^1_{\mathcal A}(U)$ is defined by,
\begin{equation}\label{H1-A}
\mathcal H_{\mathcal A}^1(U)=\{u\in L^2(U);\quad \nabla_{\mathcal
A}u\in L^2(U)\}.
\end{equation}
By Friedrich's Theorem, the quadratic form (\ref{qf}) defines a
self-adjoint operator\footnote{This is a linear elliptic operator
with discontinuous coefficients. The theory of such operators is
well treated, see \cite{Tr} for example.}, the bottom of its
spectrum is given by~:
\begin{equation}\label{EV}
\mu^{(1)}(\kappa,H)=\inf_{\phi\in\mathcal H_{\kappa H\mathbf
F}^1(\mathbb R^2),\phi\not=0}\left(\frac{\mathcal
Q[\kappa,H](\phi)}{\|\phi\|_{L^2(\mathbb R^2)}^2}\right).
\end{equation}
In terms of (\ref{EV}), we define the following `local' upper critical field,
\begin{equation}\label{HC3-loc}
H_{C_3}(a,m;\kappa)=\inf\{H>0;\quad \mu^{(1)}(\kappa,H)\geq 0\}.
\end{equation}
Below $H_{C_3}$, normal states will loose their stability. Our aim
is to estimate $H_{C_3}$ as $\kappa\to+\infty$. It could be more
convenient to use the notation $H_{C_3}^{\rm loc}$ rather than that
in (\ref{HC3-loc}). However, there are many reasonable definitions
of the upper critical field all of which are proved to coincide for
a `standard' model and when the Ginzburg-Landau parameter $\kappa$
is sufficiently large (cf. \cite{FoHe06, FoHe}). We have chosen the
definition of the upper critical field in (\ref{HC3-loc}) because
its analysis is actually purely spectral. We hope to verify in the
near future that even for this model, other definitions of the upper
critical field will coincide with the definition given in
(\ref{HC3-loc}).\\ 

We state now our main results.

\begin{thm}\label{k2Th1}
There exists a function
$\alpha:\,]0,+\infty[\times]0,+\infty[\,\mapsto\,]\Theta_0,1]$ such
that, given $a,m>0$, the upper critical field satisfies,
\begin{equation}\label{k2Th1-eq}
H_{C_3}(a, m;\kappa)=\frac{\kappa}{\alpha(a,
m)}\left(1+o(1)\right),\quad (\kappa\to+\infty).
\end{equation}
Here $\Theta_0\in]0,1[$ is a universal constant.
\end{thm}

Compared with the former literature (\cite[and references
therein]{HePa, LuPa1}), we observe that the value of the upper
critical field can be strongly modified. It is well known to
physicists that the upper critical field for type~II superconductors
is strongly dependent on the type of the
material placed adjacent to the superconductor (cf.~\cite{Hu}).\\
To prove Theorem~\ref{k2Th1}, we need various estimates on the
bottom of the spectrum $\mu^{(1)}(\kappa,H)$ in the regime
$\kappa,H\to+\infty$ (cf. (\ref{EV})). Eigenvalue asymptotics for
linear elliptic operators with discontinuous coefficients arise in
other contexts (cf.~\cite{Ji}), but here the problem is different.
We follow the technique of Helffer-Morame~\cite{HeMo3} by analyzing
the model case when $\Omega=\mathbb R\times\mathbb R_+$ is the
half-plane. Actually, let us consider the quadratic form,
\begin{equation}\label{k2HP-qf}
\mathcal H_{\kappa H\mathcal A_0}^1(\mathbb R^2)\ni \phi\mapsto
\mathcal Q_{\mathbb R\times\mathbb R_+}[\kappa,H](\phi),
\end{equation}
where
\begin{eqnarray*}
&&\mathcal Q_{\mathbb R\times\mathbb R_+}[\kappa,H](\phi)=
\int_{\mathbb R\times\mathbb R_+}\left(|(\nabla_{\kappa H
A_0}\phi|^2-\kappa^2|\phi|^2\right)dx+\int_{\mathbb R\times\mathbb
R_-}\left(\frac1{m}|\nabla_{\kappa H
A_0}\phi|^2+a\kappa^2|\phi|^2\right)dx,\end{eqnarray*} and the
magnetic potential $A_0$ is defined by~:
\begin{equation}\label{A0}
A_0(x_1,x_2)= (-x_2,0),\quad\forall (x_1,x_2)\in\mathbb
R\times\mathbb R.
\end{equation}
We denote by
\begin{equation}\label{II-EV-HP}
\mu^{(1)}(\kappa,H;\mathbb R\times\mathbb R_+)=\inf_{\phi\in\mathcal
H_{\kappa H\mathcal A_0}^1(\mathbb R^2),\phi\not=0}\frac{\mathcal
Q_{\mathbb R\times\mathbb
R_+}[\kappa,H](\phi)}{\|\phi\|^2_{L^2(\mathbb R^2)}}.
\end{equation}
Performing the scaling $t=(\kappa H)^{1/2}x_2$ and $z=(\kappa
H)^{1/2}x_1$, we get with $\alpha=\kappa/H$,
\begin{equation}\label{k2HP}
\mu^{(1)}(\kappa,H;\mathbb R\times\mathbb R_+)=\kappa
H\,\beta(a,m,\alpha). \end{equation} Here $\beta(a,m,\alpha)$ is the
bottom of the spectrum of the self-adjoint operator associated to
the quadratic form
$$\mathcal H_{A_0}^1(\mathbb R^2)\ni v\mapsto
\mathcal Q[a, m,\alpha](v),$$ which is defined by,
\begin{eqnarray}\label{II-scaledQF}
\mathcal Q[a,m,\alpha](v)&=& \int_{t>0} \left(|\partial_t
v|^2+|(t-i\partial_z)v|^2-\alpha|v|^2\right)dt dz\\
&&+\int_{t<0}\left(\frac{1}{m}\left[|\partial_t
v|^2+|(t-i\partial_z)v|^2\right]+a\alpha|v|^2\right)d\tau
dz.\nonumber
\end{eqnarray}
We define $\alpha(a,m)$ as the solution, which will be shown to
exist uniquely in Theorem~\ref{IIestalp-rem}, of the equation
$\beta(a,m,\alpha)=0$. Notice that this will correspond to the
magnetic field $H=\kappa/[\alpha(a,m)]$  that satisfies
$\mu^{(1)}(\kappa,H;\mathbb R\times\mathbb R_+)=0$.\\

In the next theorem, we describe the behavior of the function
$\alpha$.

\begin{thm}\label{k2Th2}
Given $a>0$, the function $m\mapsto \alpha(a,m)$ is strictly
decreasing, $\alpha(a,m)=1$ if $m\leq 1$, $\Theta_0<\alpha(a,m)<1$ if $m>1$, and
\begin{equation}\label{k2Th2-eq1}
\lim_{m\to1^+}\alpha(a,m)=1.
\end{equation}
Moreover, $\alpha(a,m)$ has the following asymptotic expansion as
$m$ tends to $\infty$~:
\begin{equation}\label{k2Th2-eq2}
\alpha(a,m)=\Theta_0+\frac{3\sqrt{a\Theta_0}C_1}{\sqrt{m}}+\mathcal
O\left(\frac1{m}\right),
\end{equation}
where $C_1>0$ is a universal constant.
\end{thm}

Let us mention that the universal constants $\Theta_0$ and $C_1$ are
defined via auxiliary spectral problems (cf.~(\ref{Th-0}) and
(\ref{C1})) and were already present in the analysis of the
`standard model', see however \cite{BaPhTa, BeSt,  HePa, Kach,
LuPa1} ($\Theta_0$ is indeed the bottom of the spectrum of the
Neumann realization of the Schr\"odinger operator with constant
magnetic field in $\mathbb R\times\mathbb R_+$).
Numerically~\cite{StdeGe, BoHe}, one finds that $\Theta_0\sim0.59$.
Theorem~\ref{k2Th2} proves the validity regime of the results of
\cite{Hu}.\\

For sufficiently large values of the parameter $m$, we are able to
obtain a two-term asymptotic expansion of the upper critical field,
where the scalar curvature plays a major role.

\begin{thm}\label{k2Th3}
Given $a>0$, there exist a constant ${m}_0>1$ and a function
$$\mathcal C_1(a,\cdot):\,[ m_0,+\infty[\,\mapsto\mathbb R_+$$
such that, if $m>0$ verifies $m\geq m_0$, then the upper critical
field satisfies,
\begin{equation}\label{k2Th3-eq1}
H_{C_3}(a, m;\kappa)=\frac{\kappa}{\alpha\left(a,m\right)}
+\frac{\mathcal
C_1\left(a,m\right)}{\alpha\left(a,m\right)^{3/2}}(\kappa_{\rm
r})_{\max}+\mathcal O\left(\kappa^{-1/3}\right),\quad \text{\rm as
}\kappa\to+\infty,
\end{equation}
where $\kappa_{\rm r}$ denotes the scalar curvature of
$\partial\Omega$.
\end{thm}

Let us explain what stands behind the statement of
Theorem~\ref{k2Th3}. As in \cite{LuPa2000}, we look for a {\it
formal eigenfunction} corresponding to the eigenvalue
(\ref{II-EV-HP}) in the form~:
\begin{equation}\label{EF-1/2plane}
v(x_1,x_2)=\exp\left(-i\zeta_0 (\kappa
H)^{1/2}x_1\right)f\left((\kappa H)^{1/2} x_2\right), \quad
(x_1,x_2)\in\mathbb R\times\mathbb R,\end{equation} that is, we ask
for $\zeta_0\in \mathbb R$ and $f\in H^1(\mathbb R)\cap L^2(\mathbb
R;\tau^2 d\tau)$ such that
\begin{eqnarray*}&&\hskip-1.5cm\int_{t>0}
\left(|\partial_t
f|^2+|(t+\zeta_0)f|^2-\alpha|f|^2\right)d\tau \\
&&\hskip1.5cm+\int_{t<0}\left(\frac{1}{m}\left[|\partial_t
f|^2+|(t+\zeta_0)f|^2\right]+a\alpha|f|^2\right)d\tau=0.
\end{eqnarray*}
The existence of $\zeta_0$ occurs only when $m>1$ (cf.
Theorem~\ref{min-thm1}), and, unlike the case of \cite{HeMo3,
LuPa2000}, the uniqueness of $\zeta_0$ is not obvious. We are able
to obtain uniqueness of $\zeta_0$ only in the regime $m\to+\infty$,
which explains why the conclusion of Theorem~\ref{k2Th3} is limited
to sufficiently large values of $m$. However, we believe that that
this restriction is technical and we conjecture that the function
$\mathcal C_1(a,\cdot)$ can be extended to $]1,m_0]$ so that the
asymptotic expansion (\ref{k2Th3-eq1}) will hold for any $m>1$.

\begin{rem}\label{k2Re3}\
\begin{enumerate}
\item In the regime $m\to+\infty$,
Theorems~\ref{k2Th1} and \ref{k2Th2} give,  to a first order
approximation, the same behavior as in~\cite{LuPa1}. This was
predicted by the formal computations of~\cite{Chetal}.
\item Let us note that the function $\mathcal C_1(\cdot,\cdot)$
satisfies\footnote{This will follow from Proposition~\ref{II-M1M3},
the expression of $\mathcal C_1$ (cf.~(\ref{C1-Fsec}) and
(\ref{II-C1a,m})) and the asymptotic behavior as $m\to+\infty$
(cf.~Proposition~\ref{lim1+,+infty}).}
$$\lim_{m\to+\infty}\mathcal C_1(a,m)=(1+6a\Theta_0^2)C_1,$$ so we
recover, for $a=0$ and $m=+\infty$  in (\ref{k2Th3-eq1}), the
two-term asymptotic expansion of Helffer-Pan~\cite{HePa}.
\end{enumerate}
\end{rem}

\begin{rem}\
\begin{enumerate}
\item
It would be desirable to remove the hypothesis of smoothness of the
boundary $\partial\Omega$. As in~\cite{Bon, BoFo}, one can perhaps
consider  a piecewise smooth domain.
\item After having obtained Theorems~\ref{k2Th1}-\ref{k2Th3},
we have learned  about the existence of \cite{GiJa}, where the
authors deal only with the  one-dimensional case.
\end{enumerate}
\end{rem}

\subsection{Comparison with the de\,Gennes model}
The physicist de\,Gennes~\cite{deGe} proposes to model the proximity
effect by means of a `Robin type' boundary condition. He assumes
that the order parameter satisfies,
\begin{equation}\label{k2Re1-eq1}
\nu\cdot(\nabla-i\kappa HA)\psi+\tilde\gamma(\kappa;x)\psi=0\quad
\text{on }\partial\Omega,
\end{equation}
and that this condition permits one to ignore the behavior of $\psi$
outside $\Omega$. The function $\tilde\gamma(\kappa;\cdot)$ is
supposed to be smooth and is called (when it is constant) the
de\,Gennes parameter. In this case, one has to replace the energy
(\ref{E-T}) by~:
\begin{equation}\label{k2Re1-eq2}
H^1(\Omega;\mathbb C)\times H^1(\Omega;\mathbb
R^2)\ni(\psi,A)\mapsto\mathcal E(\psi,A)=\mathcal
G_s(\psi,A)+\int_{\partial\Omega}\tilde\gamma(\kappa;x)|\psi(x)|^2d
\mu_{|_{\partial\Omega}}(x),
\end{equation}
where $\mathcal G_s$ is defined by~:
\begin{equation}\label{E-s}
\mathcal G_s(\psi,A)=\int_\Omega \left\{|\nabla_{\kappa
HA}\psi|^2-\kappa ^2|\psi|^2+\frac{\kappa^2}2|\psi|^4+(\kappa
H)^2|{\rm curl}\,A-1|^2\right\}dx.
\end{equation}
Let us take $\tilde\gamma(\kappa;x)=\kappa^\delta\gamma_0$ with
$\delta\geq0$ and $\gamma_0\in\mathbb R$. We can define an upper
critical field $H_{C_3}(\delta,\gamma_0;\kappa)$ as before (i.e.
below $H_{C_3}$, normal states will loose their stability). In the
next theorem, we give a first order approximation of
$H_{C_3}(\delta,\gamma_0;\kappa)$.

\begin{thm}\label{k2Re1}
There exists a strictly increasing function $\mathbb
R\ni\gamma\mapsto\Theta(\gamma)$ such that, as $\kappa\to+\infty$,
we have the following asymptotics~:
\begin{equation}\label{k2Re1-eq3}
H_{C_3}(\delta,\gamma_0;\kappa)=\left\{
\begin{array}{cl}
\displaystyle\frac{\kappa}{\Theta_0}(1+o(1)),&\text{\rm if }0\leq\delta<1\text{ \rm or if }\gamma_0=0;\\
&\\
\displaystyle\frac{\kappa}{\Theta(\gamma_0\cdot\ell(\gamma_0))}(1+o(1)),&
\text{\rm
if }\delta=1;\\
&\\
\kappa,&\text{\rm if }\delta>1 \text{ \rm and }\gamma_0>0;\\
&\\
\left(\displaystyle\frac{\gamma_0}{\eta_0}\right)^2\kappa^{2\delta-1},&
\text{\rm if }\delta>1\text{ \rm and }\gamma_0<0.
\end{array}\right.
\end{equation}
Here $\ell(\gamma_0)$ is the unique solution of the implicit
equation
$$\Theta(\gamma_0\cdot\ell(\gamma_0))=\ell(\gamma_0)^2$$
and $\eta_0$ is the unique zero of $\Theta(\cdot)$.
\end{thm}

For a precise definition of the function $\Theta(\cdot)$, see
(\ref{Th-gam}). In particular, we have $\Theta(0)=\Theta_0$. The
existence of the function $\ell(\cdot)$ is proved in
Lemma~\ref{app-pr}.

\begin{rem}\label{k2Re2}
Let us consider the case of Theorem~\ref{k2Th1} when $m>1$. In this
case, $\alpha\left(a,m\right)\in]\Theta_0,1[$ (cf.
Theorem~\ref{k2Th2}), and there exists a unique
$\gamma\left(a,m\right)>0$ such that
$$\alpha\left(a,m\right)=\Theta\left(
\gamma\left(a,m\right)\right).$$ Then by putting,
$$\gamma_0=\frac{\gamma\left(a,m\right)}
{\sqrt{\alpha\left(a,m\right)}},\quad
\ell(\gamma_0)=\sqrt{\alpha\left(a,m\right)},$$ we get,
$$H_{C_3}(a,m;\kappa)=\frac{\kappa}{\Theta(\gamma_0\cdot\ell(\gamma_0))}\left(1+o(1)\right),\quad
as\quad \kappa\to+\infty.$$ Therefore, the result of
Theorem~\ref{k2Th1}  corresponds to the following boundary condition
$$\nu\cdot(\nabla-i\kappa HA)\psi+\kappa\gamma_0\psi=0,\quad\textrm{ on }\partial\Omega.$$
\end{rem}

\begin{rem}\label{ext-thm}
In Theorem~\ref{k2Re1}, if $0\leq\delta\leq1$, one can still obtain
an asymptotic expansion of $H_{C_3}(\delta,\gamma_0;\kappa)$
involving the scalar curvature (cf.~\cite{HePa, Kach1}).
\end{rem}

Remark~\ref{k2Re2} suggests that one can replace the spectral
problem (\ref{qf})-(\ref{EV}) by a suitable problem in $\Omega$
(with some de\,Gennes boundary condition) having the same ground
state energies and whose ground states coincide in $\Omega$.
However, as the case of the half-plane model will show, this will
not be the case and the
best one can hope is the convergence of the ground state energies.\\
Actually, if this were the case for the half-plane, i.e. there
exists $\zeta_0=\zeta_0(a,m)\in\mathbb R$ such that the function $v$
given by (\ref{EF-1/2plane}) is an eigenfunction of
(\ref{k2HP-qf})-(\ref{II-EV-HP}) and satisfies the boundary
condition:
$$\partial_{x_2}v=\gamma_0(a,m)\,v\quad{\rm on}~\partial\mathbb
R\times\mathbb R_+,$$ then we get that
$\zeta_0(a,m)=\xi\big{(}\gamma(a,m)\big{)}$, where $\gamma_0(a,m)$
and $\gamma(a,m)$ are given by Remark~\ref{k2Re2}, and $\xi(\cdot)$
will be defined in (\ref{xi(gam)}). On the other hand, by the
discussion in Subsection~\ref{subSec-m->infty}, and in particular
Proposition~\ref{lim+infty}, we get that
$$\lim_{m\to+\infty}\sqrt{m}\bigg{(}\zeta_0(a,m)-\xi\big{(}\gamma(a,m)\big{)}\bigg{)}>0,$$
which shows that it is impossible for the ground state to satisfy
the boundary condition corresponding to the right value of the
ground state energy.\\
One should also mention in this direction the result for the
nonlinear problem  obtained recently in \cite{kach06}, where the
author proves that for the case without magnetic field, $H=0$, all
minimizers of (\ref{E-T}) are gauge equivalent to a real phase
$(u_\kappa,0)$ that satisfies as $\kappa\to+\infty$~:
$$\nu\cdot\nabla
u_\kappa=\kappa\sqrt{\frac{a}{2m}}\,u_\kappa(1+o(1))\quad{\rm
on}~\partial\Omega.$$ When compared with
Theorems~\ref{k2Th2},~\ref{k2Re1} and Remark~\ref{k2Re2} we get,
especially  for $m=1$, that for each regime of the applied magnetic
field one should associate a different de\,Gennes boundary
condition. The physical interpretation is that the penetration
length, which de\,Gennes measures in terms of his parameter in the
boundary condition, is strongly dependent on the applied magnetic
field (as predicted in \cite{Pa}).

\subsection{Organization of the paper}
In Section~\ref{Aux-mat}, we recall auxiliary material that we shall
use frequently in the paper. The analysis of the model problem
(\ref{k2HP-qf}) leads in Section~\ref{k2-MP} to the spectral
analysis of a family of ordinary differential operators. We obtain
by elementary arguments most of the properties announced in
Theorem~\ref{k2Th2}, and we complete its proof by a fine asymptotic
analysis when the parameter
$m$ is large.\\
In Section~\ref{Sec-MF}, we analyze a `refined' family of model
operators, whose study is essential for the proof of
Theorem~\ref{k2Th3}.\\
In Section~\ref{k2-Est-btSp}, we establish a simpler formula of the
upper critical field. Using the results of Section~\ref{k2-MP}, we
are able to follow a similar analysis to \cite{HeMo3} for estimating
the bottom of the spectrum $\mu^{(1)}(\kappa,H)$ and we give a proof
of
Theorem~\ref{k2Th1}.\\
Since the variational problem (\ref{EV}) is over the whole plane
$\mathbb R^2$, minimizers do not always exist. In
Section~\ref{Sec-Pe}, we establish using  Persson's Lemma~\cite{Pe}
the existence of minimizers to (\ref{EV}) when the intensity of the
magnetic field $H$ is near $H_{C_3}$. We prove also by using the
technique of Agmon estimates~\cite{Ag} that the minimizers of
(\ref{EV}) decay exponentially fast away from the boundary
$\partial\Omega$.\\
We are now ready in Section~\ref{k2-2Tasy} to imitate the analysis
of Helffer-Morame~\cite[Section~11]{HeMo3} and to derive a two
term-asymptotic expansion of $\mu^{(1)}(\kappa,H)$ and we use it to
prove Theorem~\ref{k2Th3}.\\
Finally, in Appendix~\ref{App-A}, we give  a proof for the
asymptotics announced in Theorem~\ref{k2Re1}.

\section{Auxiliary material}\label{Aux-mat}
\subsection{A family of ordinary differential operators.}\label{ODE}
The analysis of a canonical operator in the half-plane $\mathbb
R\times\mathbb R_+$ with de\,Gennes boundary condition leads us
naturally to a family of ordinary differential operators (cf.
\cite{Kach1}). Given $(\gamma,\xi)\in\mathbb R\times\mathbb R$, we
define the quadratic form,
\begin{equation}\label{k2-qfk1}
B^1(\mathbb R_+)\ni u\mapsto q[\gamma,\xi](u)=\int_{\mathbb
R_+}\left(|u'(t)|^2+|(t-\xi)u(t)|^2\right)dt+\gamma|u(0)|^2,
\end{equation}
where, for a positive integer $k\in\mathbb N$ and a given interval
$I\subseteq\mathbb R$, the space $B^k(I)$ is defined by~:
\begin{equation}\label{Bk-sp}
B^k(I)=\{u\in H^k(I);\quad t^ju(t)\in L^2(I),\quad \forall
j=1,\cdots, k\}.
\end{equation}
By Friedrichs' Theorem, we can associate to the quadratic form
(\ref{k2-qfk1}) a self adjoint operator  $\mathcal L[\gamma,\xi]$
with domain,
$$D(\mathcal L[\gamma,\xi])=\{u\in B^2(\mathbb R_+);\quad
u'(0)=\gamma u(0)\},$$ and associated to the differential operator,
\begin{equation}\label{L-dop}
\mathcal L[\gamma,\xi]=-\partial_t^2+(t-\xi)^2.
\end{equation} We denote by
$\{\lambda_j(\gamma,\xi)\}_{j=1}^{+\infty}$ the increasing sequence
of eigenvalues of $\mathcal L[\gamma,\xi]$. When $\gamma=0$ we
write,
\begin{equation}\label{l-ga=0}
\lambda_j^N(\xi):=\lambda_j(0,\xi),\quad \forall j\in\mathbb N,\quad
\mathcal L^N[\xi]:=\mathcal L[0,\xi].
\end{equation}
We also denote by $\{\lambda^D_j(\xi)\}_{j=1}^{+\infty}$ the
increasing sequence of eigenvalues of the Dirichlet realization of
$-\partial_t^2+(t-\xi)^2$.\\
By the min-max principle, we have,
\begin{equation}\label{l1-ga,xi}
\lambda_1(\gamma,\xi)=\inf_{u\in B^1(\mathbb
R_+),u\not=0}\frac{q[\gamma,\xi](u)}{\|u\|^2_{L^2(\mathbb R_+)}}.
\end{equation}
Let us denote by $\varphi_{\gamma,\xi}$ the positive (and
$L^2$-normalized) first eigenfunction of $\mathcal L[\gamma,\xi]$.
It is proved in \cite{Kach1} that the functions
$$(\gamma,\xi)\mapsto\lambda_1(\gamma,\xi),\quad (\gamma,\xi)\mapsto
\varphi_{\gamma,\xi}\in L^2(\mathbb R_+)$$ are regular (i.e. of
class $C^\infty$), and we have the following formulas,
\begin{eqnarray}
&&\partial_\xi\lambda_1(\gamma,\xi)=-\left(\lambda_1(\gamma,\xi)-\xi^2+\gamma^2\right)
|\varphi_{\gamma,\xi}(0)|^2,\label{Th7.5.5}\\
&&\partial_\gamma\lambda_1(\gamma,\xi)=|\varphi_{\gamma,\xi}(0)|^2.\label{Th7.5.6}
\end{eqnarray}
Notice that (\ref{Th7.5.6}) will yield that the function
$$(\gamma,\xi)\mapsto\varphi_{\gamma,\xi}(0)$$
is also regular of class $C^\infty$.\\
We define the function~:
\begin{equation}\label{Th-gam}
\Theta(\gamma)=\inf_{\xi\in\mathbb R}\lambda_1(\gamma,\xi).
\end{equation}
It is a result of~\cite{DaHe} that there exists a unique
$\xi(\gamma)>0$ such that,
\begin{equation}\label{xi-gam}
\Theta(\gamma)=\lambda_1(\gamma,\xi(\gamma)),
\end{equation}
and $\xi(\gamma)$ satisfies (cf. \cite{Kach1}),
\begin{equation}\label{xi(gam)}
\xi(\gamma)^2=\Theta(\gamma)+\gamma^2.
\end{equation}
Moreover, the function $\Theta(\gamma)$ is of class $C^\infty$ and
satisfies,
\begin{equation}\label{Th'-gam}
\Theta'(\gamma)=|\varphi_\gamma(0)|^2,
\end{equation}
where $\varphi_\gamma$ is the positive (and $L^2$-normalized)
eigenfunction associated to $\Theta(\gamma)$~:
\begin{equation}\label{varphi-gam}
\varphi_\gamma=\varphi_{\gamma,\xi(\gamma)}. \end{equation} When
$\gamma=0$, we write,
\begin{equation}\label{Th-0}
\Theta_0:=\Theta(0),\quad \xi_0:=\xi(0).
\end{equation}
We define also the universal constant $C_1$ by,
\begin{equation}\label{C1}
C_1:=\frac{|\varphi_0(0)|^2}3.
\end{equation}

Another important fact is the following consequence of standard
Sturm-Liouville theory.
\begin{lem}\label{HeRu}
For any $\xi\in\mathbb R$, we have,
$$\lambda_2^N(\xi)>\lambda^D_1(\xi).$$
\end{lem}

\subsection{Boundary Coordinates}\label{sbS-BC} We recall now
the definition of the standard coordinates that straightens a
portion of the boundary $\partial\Omega$. Given $t_0>0$, let us
introduce the following neighborhood of the boundary,
\begin{equation}\label{N-t0}
\mathcal N_{t_0}=\{x\in\mathbb R^2;\quad {\rm
dist}(x,\partial\Omega)<t_0\}.
\end{equation}
As the boundary is smooth, let
$s\in]-\frac{|\partial\Omega|}2,\frac{|\partial\Omega|}2]\mapsto
M(s)\in\partial\Omega$ be a regular parametrization of
$\partial\Omega$ that satisfies~:
$$\left\{
\begin{array}{l}
s \text{ is
the oriented `arc length' between }M(0) \text{ and }
M(s).\\
T(s):=M'(s) \text{ is a unit tangent vector
to }\partial\Omega\text{ at the point }M(s).\\
\text{The orientation is positive, i.e. }{\rm det}(T(s),\nu(s))=1.
\end{array}
\right. $$ We recall that $\nu(s)$ is the unit outward normal of
$\partial\Omega$ at the point $M(s)$. The scalar curvature
$\kappa_{\rm r}$ is now defined by~:
\begin{equation}\label{kappa-r}
T'(s)=\kappa_{\rm r}(s)\nu(s).
\end{equation}
When $t_0$ is sufficiently small, the map~:
\begin{equation}\label{Phi(s,t)}
\Phi:\,]-|\partial\Omega|/2,|\partial\Omega|/2]\times
]-t_0,t_0[\,\ni(s,t)\mapsto M(s)-t\nu(s)\in \mathcal N_{t_0},
\end{equation}
is a diffeomorphism. For $x\in\mathcal N_{t_0}$, we write,
\begin{equation}\label{Phi-1}
\Phi^{-1}(x):=(s(x),t(x)),
\end{equation}
where
$$t(x)={\rm dist}(x,\partial\Omega)\text{ if }x\in\Omega \quad
\text{and }t(x)=-{\rm dist}(x,\partial\Omega)\text{ if
}x\not\in\Omega.$$ The jacobian of the transformation $\Phi^{-1}$ is
equal to,
\begin{equation}\label{Jac}
a(s,t)={\rm det}\left(D\Phi^{-1}\right)=1-t\kappa_{\rm r}(s).
\end{equation}
To a vector field $A=(A_1,A_2)\in H^1(\mathbb R^2;\mathbb R^2)$, we
associate the vector field
$$\tilde A=(\tilde A_1,\tilde A_2)\in
H^1(]-|\partial\Omega|/2,|\partial\Omega|/2]\times]-t_0,t_0[;\mathbb
R^2)$$
by the following relations~:
\begin{equation}\label{chmnew}
\tilde A_1(s,t)=(1-t\kappa_{\rm r}(s)) \vec A(\Phi(s,t))\cdot
M'(s),\quad \tilde A_2(s,t)=\vec A(\Phi(s,t))\cdot\nu(s).
\end{equation}
We get then the following change of variable formulas.
\begin{proposition}\label{transf}
Let $u\in H^1_A(\mathbb R^2)$ be supported in $\mathcal N_{t_0}$.
Writing $\widetilde u(s,t)=u(\Phi(s,t))$, then we have~:
\begin{equation}\label{qfstco}
\int_{\Omega}\left|(\nabla-iA)u\right|^2dx=
\int_{-\frac{|\partial\Omega|}2}^{\frac{|\partial\Omega|}2}
\int_{0}^{t_0}\left[ |(\partial_s-i\tilde A_1)\widetilde
u|^2+a^{-2}|(\partial_t-i\tilde A_2)\widetilde u|^2\right]a\,dsdt,
\end{equation}
\begin{equation}\label{qfstco'}
\int_{\Omega^c}\left|(\nabla-iA)u\right|^2dx=
\int_{-\frac{|\partial\Omega|}2}^{\frac{|\partial\Omega|}2}
\int_{-t_0}^{0}\left[ |(\partial_s-i\tilde A_1)\widetilde
u|^2+a^{-2}|(\partial_t-i\tilde A_2)\widetilde u|^2\right]a\,dsdt,
\end{equation}
and
\begin{equation}\label{nostco}
\int_{\mathbb R^2}
|u(x)|^2\,dx=\int_{-\frac{|\partial\Omega|}2}^{\frac{|\partial\Omega|}2}
\int_{-t_0}^{t_0} |\widetilde u(s,t)|^2a\,dsdt.
\end{equation}
\end{proposition}

We have also the relation~:
$$\left(\partial_{x_1}A_2-\partial_{x_2}A_1\right)dx_1\wedge dx_2=
\left(\partial_s\tilde A_2-\partial_t\tilde A_1\right)a^{-1}ds\wedge
dt,$$ which gives,
$${\rm curl}_{(x_1,x_2)}\, A=\left(1-t\kappa_{\rm r}(s)\right)^{-1}{\rm curl}_{(s,t)}\,\tilde A.$$

We give in the next proposition a standard choice of gauge.

\begin{prop}\label{Agd1}
Consider a vector field $A=(A_1,A_2)\in C^1_{\rm loc}(\mathbb
R^2;\mathbb R^2)$ such that
$${\rm curl }\,A=1\quad \text{ in }\mathbb R^2.$$ For each point $x_0\in\partial\Omega$, there
exist a neighborhood $\mathcal V_{x_0}\subset\mathcal N_{t_0}$ of
$x_0$ and a smooth real-valued function $\phi_{x_0}$ such that the
vector field $A_{new}:=A-\nabla\phi_{x_0}$ satisfies in $\mathcal
V_{x_0}$~:
\begin{equation}\label{HeMo'}
\tilde A_{new}^2=0,
\end{equation}
and,
\begin{equation}\label{HeMo}
\tilde A_{new}^1=-t\left(1-\frac t2\kappa_{\rm r}(s)\right),
\end{equation}
with $\tilde A_{new}=(\tilde A_{new}^1,\tilde A_{new}^2)$.
\end{prop}
\subsection{The Neumann and Dirichlet magnetic Schr\"odinger
operators} Let us consider the differential operator,
$$\mathcal P_\varepsilon=-(\nabla-\varepsilon^{-2}\mathbf F)^2,$$
where $\varepsilon$ is a small parameter. Given a domain
$U\subset\mathbb R^2$, we denote by $\mathcal P_{\varepsilon,U}^N$
and $\mathcal P_{\varepsilon,U}^D$ the Neumann and Dirichlet
realizations of $\mathcal P_\varepsilon$ in $U$ respectively. Then
we introduce~:
\begin{equation}\label{mu-N,D}
\mu^N(\varepsilon;U)=\inf\,{\rm Sp}\left(\mathcal
P_{\varepsilon,U}^N\right),\quad \mu^D(\varepsilon;U)=\inf\,{\rm
Sp}\left(\mathcal P_{\varepsilon,U}^D\right).
\end{equation}
We recall the following result of \cite{HeMo3}.
\begin{prop}\label{HeMo3-est-ND}
Given a domain $U\subset\mathbb R^2$ with compact and smooth
boundary, there exist constants $C,\varepsilon_0>0$ such that, if
${\rm curl}\,\mathbf F=1$ in $U$, then we have for any $\varepsilon\in]0,\varepsilon_0]$,
\begin{eqnarray}\label{HeMo3-ND-eq1}
&&\varepsilon^{-2}\leq\mu^D(\varepsilon;U)\leq \varepsilon^{-2}+
C\varepsilon^{-2}\exp\left(-\frac1\varepsilon\right),\\
&&\left|\mu^N(\varepsilon;U)-\Theta_0\varepsilon^{-2}\right|\leq
C\varepsilon^{-1}.\label{HeMo3-ND-eq2}
\end{eqnarray}
\end{prop}

We define the quadratic form,
\begin{equation}\label{qf-N}
\mathcal H_{\varepsilon^{-2}\mathbf F}^1(U)\ni u\mapsto
q_{\varepsilon,\mathbf F,U}(u)=\|(\nabla-i\varepsilon^{-2}\mathbf
F)u\|_{L^2(U)}^2.
\end{equation}

\begin{prop}\label{HeMo3-Bprop}
Suppose that $U$ has a smooth  compact boundary and that ${\rm
curl}\,\mathbf F=1$ in $U$.
\begin{enumerate}
\item
If $u\in C_0^\infty(U)$, then
\begin{equation}\label{HeMo3-int}
q_{\varepsilon,\mathbf F,U}(u)\geq \varepsilon^{-2}\|u\|_{L^2(U)}^2.
\end{equation}
\item
Given a point $z_0\in\partial U$, suppose that there exists a
function $\phi_0\in C^2(U)$ and a constant $C_0$ that depends only
on $U$ such that, upon putting,
$$\mathbf F_{new}=\mathbf F+\nabla\phi_0,$$
we have in boundary coordinates,
$$\left|\widetilde {\mathbf F}_{new}(s,t)-A_0(s,t)\right|\leq C_0
t^2,\quad\text{ in a neighborhood of }z_0,$$ where
$\widetilde{\mathbf F}_{new}$ is associated to $\mathbf F_{new}$ by
the relation (\ref{chmnew}), and the magnetic potential $A_0$ is
defined by
$$A_0(s,t)=(-t,0),\quad \forall (s,t)\in\mathbb
R\times\mathbb R_+.$$ Then  there exist constants
$C,\varepsilon_0>0$ depending only on $U$, such that, for all
$$\zeta_0,\rho,\theta>0,\quad\varepsilon\in]0,\varepsilon_0],\quad u\in\mathcal
H_{\varepsilon^{-2}\mathbf F}^1(U),\quad{\rm supp}\,u\subset
D(z_0,\zeta_0\varepsilon^\rho),$$ we have,
\begin{eqnarray}\label{HeMo3-bnd}
q_{\varepsilon,\mathbf F,U}(u)&\geq&
\left(1-C\zeta_0\varepsilon^\rho-C\varepsilon^{2\theta}\right)
q_{\varepsilon,A_0,\mathbb R\times\mathbb
R_+}\left(\exp\left(-i\frac{\phi_0}{\varepsilon^2}\right)\widetilde
u\right)\\
&&-C\varepsilon^{4\rho-2\theta-4}\|\widetilde u\|_{L^2(\mathbb
R\times\mathbb R_+)}^2,\nonumber
\end{eqnarray}
where the function $\widetilde u$ is defined by means of the change
of variables (cf. (\ref{Phi(s,t)})),
$$\widetilde u(s,t)=u\left(\Phi(s,t)\right).$$
\end{enumerate}
\end{prop}

The lower bound (\ref{HeMo3-int}) is well-known (cf.~\cite{AvHeSi})
and is  easy to prove in our case. The estimate (\ref{HeMo3-bnd}) is
essentially obtained in~\cite[p.~16]{HeMo3}. The proof consists
actually of writing the quadratic form $q_{\varepsilon,\mathbf F,U}$
in the coordinate system $(s,t)$, then doing a gauge transformation
that permits one to work with $\mathbf F_{new}$ which can be
approximated by the `canonical' magnetic potential $A_0$, and
finally of applying a Cauchy-Schwarz inequality.

Finally we recall the definition of a useful `scaled' partition of
unity of $\mathbb R^2$ attached to a covering by balls of radius
$\zeta_0\varepsilon^\rho$.
\begin{prop}\label{P-O-U}
Let $0\leq\rho\leq2$ and $\varepsilon_0>0$. There exist a constant
$C>0$ and a partition of unity $\chi_j$ of $\mathbb R^2$ such that
with,
$$\chi_j^{\varepsilon}(x):=\chi_j\left(\frac{x}{\zeta_0\varepsilon^\rho}\right),\quad
\varepsilon\in]0,\varepsilon_0]\quad {\rm and}\,\zeta_0>0,$$
we have,
\begin{eqnarray}
&&\sum_{j}|\chi_j^\varepsilon|^2=1,\label{P-O-U-eq1}\\
&&\sum_{j}|\nabla\chi_j^\varepsilon|^2\leq
C\zeta_0^{-2}\varepsilon^{-2\rho},\label{P-O-U-eq2}\\
&&{\rm supp}\,\chi_j^\varepsilon\subset
D(z_j^\varepsilon,\zeta_0\varepsilon^\rho)\text{ \rm and } \left\{
\begin{array}{l}
\text{ \rm either }{\rm supp}\,\chi_j^\varepsilon\cap
\partial \Omega=\emptyset,\\
\text{ \rm or }z_j^\varepsilon\in\partial \Omega.
\end{array}
\right.\label{P-O-U-eq3}
\end{eqnarray}
Moreover, we have the following decomposition formula,
\begin{equation}\label{IMS-f}
\forall u\in \mathcal H_{\varepsilon^{-2}\mathbf F}^1(U),\quad
q_{\varepsilon,\mathbf F,U}(u)=\sum_jq_{\varepsilon,\mathbf
F,U}\left(\chi_j^\varepsilon
u\right)-\sum_j\left\|\,|\nabla\chi_j^\varepsilon|u\right\|^2_{L^2(U)},
\end{equation}
where $U$ is either $\Omega$ or $\overline{\Omega}^c$.
\end{prop}

Formula (\ref{IMS-f}) is called in other contexts the IMS formula,
see however~\cite{CFKS}.

\section{Analysis of the canonical `interface' operator}\label{k2-MP}
The analysis of the half-plane model operator associated with the
quadratic form (\ref{k2HP-qf}) leads us to the analysis of a family
of ordinary differential operators (This is by performing a partial
Fourier transformation with respect to the second variable $z$).
\subsection{Notations and preliminaries}
Given $a,m,\alpha>0$ and $\xi\in\mathbb R$, let us define
the quadratic form~:
\begin{equation}\label{II-qf-xi}
B^1(\mathbb R)\ni u\mapsto Q[a,m,\alpha;\xi](u),
\end{equation}
where~:
\begin{eqnarray}\label{IIfq}
&&Q[a,m,\alpha;\xi](u)=\int_{\mathbb
R_+}\left(|u'(t)|^2+|(t-\xi)u(t)|^2-\alpha|u(t)|^2\right)dt\\
&&\hskip3cm+\int_{\mathbb
R_-}\left(\frac1m\left[|u'(t)|^2+|(t-\xi)u(t)|^2\right]+a\alpha|u(t)|^2\right)dt.\nonumber
\end{eqnarray}
We denote by $H[a,m,\alpha;\xi]$ the self-adjoint operator
associated to the quadratic form (\ref{II-qf-xi}) by Friedrichs'
Theorem. The domain of $H[a,m,\alpha;\xi]$ is defined by~:
\begin{equation}\label{IIdOp}
D(H[a,m,\alpha;\xi])=\left\{u\in B^1(\mathbb R);\quad u_{|_{\mathbb
R_{\pm}}}\in B^2(\mathbb R_{\pm}),\quad u'(0_+)=\frac1m
u'(0_-)\right\},
\end{equation}
and for $u\in D(H[a,m,\alpha;\xi])$, we have,
\begin{equation}\label{II-H-}
\left(H[a,m,\alpha;\xi]u\right)(t)=\left\{\begin{array}{l}
\left[\left(-\partial_t^2+(t-\xi)^2-\alpha\right)u\right](t);\quad\text{if } t>0,\\
\\
\left[\left(\frac1m\left\{-\partial_t^2+(t-\xi)^2\right\}+a\alpha\right)u\right](t);\quad
\text{if }t<0.\end{array}\right.
\end{equation}
We denote by $\mu_1(a,m,\alpha;\xi)$ the first eigenvalue of
$H[a,m,\alpha;\xi]$ which is given by the min-max principle,
\begin{equation}\label{IIpVP}
\mu_1(a,m,\alpha;\xi)=\inf_{u\in B^1(\mathbb
R),u\not=0}\frac{Q[a,m,\alpha;\xi](u)}{\|u\|^2_{L^2(\mathbb
R)}}.
\end{equation}
The eigenvalue $\mu_1(a,m,\alpha;\xi)$ is simple and there exists a
unique strictly positive (and $L^2$-normalized) eigenfunction
$f_{\alpha,\xi}^{a,m}$. To see that $f_{\alpha,\xi}^{a,m}(0)>0$, we
suppose for a contradiction that $f_{\alpha,\xi}^{a,m}(0)=0$, then
we define the function $v\in B^1(\mathbb R)$ by~:
$$v(t)=f_{\alpha,\xi}^{a,m}(t)\quad \text{ if }t>0,\quad
v(t)=0\quad \text{ if }t\leq0.$$ So $v$ is in the form domain of
$Q[a,m,\alpha;\xi]$ and an  integration by parts yields the
equality~:
$$\frac{Q[a,m,\alpha;\xi](v)}{\|v\|_{L^2(\mathbb
R)}^2}=\mu_1(a,m,\alpha;\xi).$$ Therefore, by the min-max principle,
$v$ is an eigenfunction of $H[a,m,\alpha;\xi]$ and consequently
$v\in D(H[a,m,\alpha;\xi])$. Hence $v'(0_+)=v(0_+)=0$ and so we get
by Cauchy's Uniqueness Theorem for solutions of ordinary
differential equations
that $v\equiv0$, which is the desired contradiction.\\
We can now define the following `effective de\,Gennes parameter'~:
\begin{equation}\label{Sec2,deGe}
\gamma(a,m,\alpha;\xi):=\left(\frac{\left(f_{\alpha,\xi}^{a,m}\right)'}
{f_{\alpha,\xi}^{a,m}}\right)(0_+).
\end{equation}
Using the boundary condition satisfied by $f_{\alpha,\xi}^{a,m}$, we
get,
$$\gamma(a,m,\alpha;\xi)=\frac1m\left(\frac{\left(f_{\alpha,\xi}^{a,m}\right)'}
{f_{\alpha,\xi}^{a,\alpha}}\right)(0_-).$$ In the physical
literature~\cite{deGe1, Hu, Pa}, the parameter
$\frac{1}{\gamma(a,m,\alpha;\xi)}$ is usually called the
`extrapolation length'.\\

Let us notice that the quadratic form $Q[a,m,\alpha;\xi]$ has a
fixed domain and that, given $u$ in its form domain, the function
$$\mathbb R_+^*\times\mathbb R_+^*\times\mathbb R_+^*\times\mathbb R
\ni(a,m,\alpha;\xi)\mapsto Q[a,m,\alpha;\xi](u)$$ is analytic.
Therefore, $H[a,m,\alpha;\xi]$ is a `holomorphic operator of Type~A'
with respect to the variables $a,m,\alpha,\xi$ (cf.
\cite{ka})\footnote{In \cite{ka}, the analytic perturbation theory
is developed for operators depending on a single variable. However
it is straightforward to generalize this theory to operators
depending on `many variables' (see for example \cite{BoHe}) and the
analyticity of the eigenvalues will remain true when they are
simple.}. We then get, thanks to the simplicity of
$\mu_1(a,m,\alpha;\xi)$, that the first eigenvalue and the
corresponding normalized eigenfunction of $Q[a,m,\alpha;\xi]$ are
analytic. This yields the following regularity result, proved in
\cite[Theorem~3.1.2]{kachTh} by a different approach.

\begin{thm}\label{reg}
The functions,
$$(a,m,\alpha,\xi)\mapsto\mu_1(a,m,\alpha;\xi),\quad(a,m,\alpha,\xi)\mapsto
f_{\alpha,\xi}^{a,m}\in L^2(\mathbb R),$$ and $$(a,m,\alpha,\xi)
\mapsto\gamma(a,m,\alpha;\xi)$$ are of class $C^\infty$.
\end{thm}

\begin{rem}\label{Rel-HP}
Notice that, by a partial Fourier transformation (with respect to
the variable $z$) and by separation of variables (cf.~\cite{ReSi4}),
one has, (cf.~(\ref{k2HP})),
\begin{equation}\label{Rel-Hp-eq}
\mu^{(1)}(\kappa,H;\mathbb R\times\mathbb R_+)=\kappa H\left (
\inf_{\xi\in\mathbb
R}\mu_1\left(a,m,\frac{\kappa}{H};\xi\right)\right).
\end{equation}
\end{rem}

\subsection{Variation with respect to $\xi$}
We study in this section the variations of the function~:
$$\xi\mapsto\mu_1(a,m,\alpha;\xi).$$
It results from the min-max principle the following lemma.
\begin{lem}\label{IIestalp}
For  all $a,m,\alpha>0$ and $\xi\in\mathbb R$, the eigenvalue
$\mu_1(a,m,\alpha;\xi)$ satisfies~:
\begin{eqnarray}\label{IIestalp-eq}
&&\min\left(\lambda_1^N(\xi)-\alpha,\frac1m\lambda_1^N(-\xi)+a\alpha\right)\\
&&\leq\mu_1(a,m,\alpha;\xi)\nonumber\\
&&\leq
\min\left(\lambda_1^D(\xi)-\alpha,\frac1m\lambda_1^D(-\xi)+a\alpha\right).\nonumber
\end{eqnarray}
\end{lem}
Combining with previous well known results about
$\lambda_1^N(\cdot)$ and $\lambda_1^D(\cdot)$ (cf.~\cite{BoHe,
HeMo3}), we get that the function $\mu_1(a,m,\alpha;\cdot)$ is
bounded and,
\begin{equation}\label{lim-pm-inf}
\lim_{\xi\to-\infty}\mu_1(a,m,\alpha;\xi)=\frac1m+a\alpha,\quad
\lim_{\xi\to+\infty}\mu_1(a,m,\alpha;\xi)=1-\alpha.
\end{equation}

In the next proposition, we give an explicit formula for the
derivative of $\mu_1(a,m,\alpha;\cdot)$.

\begin{prop}\label{IIderVP}
For all $a,m,\alpha>0$ and $\xi\in\mathbb R$, we have~:
\begin{equation}\label{IIderVP-eq}
\begin{array}{l}
\partial_\xi\mu_1(a,m,\alpha;\xi)\\
=\left[(m-1)|\gamma(a,m,\alpha;\xi)|^2+ \left(1-\frac1m\right)\xi^2
-(1+a)\alpha\right] \left|f_{\alpha,\xi}^{a,m}(0)\right|^2,
\end{array}
\end{equation}
where $\gamma(a,m,\alpha;\xi)$ is defined in (\ref{Sec2,deGe}).
\end{prop}
\paragraph{\bf Proof.}
We follow the method of Dauge-Helffer~\cite{DaHe}. To simplify, we
shall omit the reference to the parameters $a,m,\alpha$ and write
$\mu_1(-;\xi)$, $f_\xi$ and $H[-;\xi]$ respectively for the first
eigenvalue $\mu_1(a,m,\alpha;\xi)$, the first eigenfunction
$f_{\alpha,\xi}^{a,m}$ and the operator
$H[a,m,\alpha;\xi]$.\\
Consider a real $\tau>0$. Notice that~:
$$H[-;\xi]f_{\xi+\tau}(t+\tau)=\mu_1(-;\xi+\tau)f_{\xi+\tau}(t+\tau),\quad
(t\not\in[-\tau,0]).
$$
An integration by parts yields~:
\begin{eqnarray}\label{IIDaHeT-}
&&\int_{-\infty}^{-\tau}H[-;\xi]f_\xi(t)\,f_{\xi+\tau}(t+\tau)\,dt=
\int_{-\infty}^{-\tau}f_\xi(t)H[-;\xi]f_{\xi+\tau}(t+\tau)\,dt\\
&&+\frac1m\left\{f_\xi(-\tau)f'_{\xi+\tau}(0_-)-f'_\xi(-\tau)f_{\xi+\tau}(0)\right\},\nonumber
\end{eqnarray}
and,
\begin{eqnarray}\label{IIDaHeT+}
&&\int_0^{+\infty}H[-;\xi]f_\xi(t)\,f_{\xi+\tau}(t+\tau)\,dt=
\int_0^{+\infty}f_\xi(t)H[-;\xi]f_{\xi+\tau}(t+\tau)\,dt\\
&&-f_\xi(0)f'_{\xi+\tau}(\tau)+f'_\xi(0_+)f_{\xi+\tau}(\tau).\nonumber
\end{eqnarray}
By taking the sum of the preceding two equalities, we get the
following identity~:
\begin{eqnarray}\label{IIDaHeS+-}
&&\left(\mu_1(-;\xi+\tau)-\mu_1(-;\xi)\right)\int_{-\infty}^{+\infty}f_{\xi}(t)
f_{\xi+t}(t+\tau)dt\\
&&=\frac1m\left\{f'_{\xi}(-\tau)f_{\xi+\tau}(0)-f_\xi(-\tau)f'_{\xi+\tau}(0_-)\right\}
\nonumber\\
&&\hskip.5cm
+f_\xi(0)f'_{\xi+\tau}(\tau)-f'_\xi(0+)f_{\xi+\tau}(\tau)\nonumber\\
&&\hskip0.5cm+\left(\mu_1(-;\xi+\tau)-\mu_1(-;\xi)\right)\int_{-\tau}^{0}f_{\xi}(t)
f_{\xi+t}(t+\tau)dt.\nonumber
\end{eqnarray}
Using the mean value theorem, we get,
\begin{eqnarray*}
&&\left|f'_\xi(-\tau)-f'_\xi(0_-)+\tau f''_\xi(0_-)\right|\leq \tau^2\|f'''_\xi\|_{L^\infty(0,1)},\\
&&\left|f_{\xi+\tau}(\tau)-f_{\xi+\tau}(0) -\tau
f'_{\xi+\tau}(0_+)\right|\leq
\tau^2\|f''_{\xi+\tau}\|_{L^\infty(0,1)}.
\end{eqnarray*}
By the boundary condition $f'_\xi(0_+)=\frac1mf'_\xi(0_-)$, we can
rewrite the preceding estimate in the form~:
\begin{eqnarray*}
&&\frac1mf'_\xi(-\tau)f_{\xi+\tau}(0)-f'_\xi(0_+)f_{\xi+\tau}(\tau)\\
&&= -\tau f'_\xi(0_+)f'_{\xi+\tau}(0_+)-\frac\tau{m}
f_{\xi+\tau}(0)f''_\xi(0_-)+\mathcal O(\tau^2).\end{eqnarray*} The
same argument yields~:
\begin{eqnarray*}
&&f_\xi(0)f'_{\xi+\tau}(\tau)-\frac1mf_\xi(-\tau)f'_{\xi+\tau}(0_-)\\
&&= m\tau f'_{\xi}(0_+)f'_{\xi+\tau}(0_+)+\tau
f_\xi(0)f''_{\xi+\tau}(0_+)+\mathcal O(\tau^2).\end{eqnarray*}
Dividing   the two sides of (\ref{IIDaHeT+}) by $\tau$ then taking
the  limit  $\tau\to0$, we get~:
\begin{equation}
\partial_\xi\mu_1(-;\xi)=\left(m-1\right)|f'_\xi(0_+)|^2
-\frac1mf_\xi(0)f''_\xi(0_-)+ f_\xi(0)f''_\xi(0_+).\end{equation}
The boundary condition and the equations satisfied by  $f_\xi(t)$ in
$\mathbb R_-$ and $\mathbb R_+$ respectively permit now to deduce
Formula (\ref{IIderVP-eq}).\hfill$\Box$\\

Let us define the set of points of minima of
$\mu_1(a,m,\alpha;\xi)$,
\begin{equation}\label{Min-xi}
{\rm M}(a,m,\alpha)=\{\eta\in\mathbb R;\quad
\mu_1(a,m,\alpha;\eta)=\inf_{\xi\in\mathbb
R}\mu_1(a,m,\alpha;\xi)\}.
\end{equation}

Formula (\ref{IIderVP-eq}) permits to obtain sufficient conditions
on $a,m,\alpha$ for the set ${\rm M}(a,m,\alpha)$ to be empty or
not.

\begin{thm}\label{min-thm1}
If $m\leq1$, then the set ${\rm M}(a,m,\alpha)$ is empty and
\begin{equation}\label{IICcr-eq}
\inf_{\xi\in\mathbb R} \mu_1(a,m,\alpha;\xi)=1-\alpha.
\end{equation}
On the other hand, let $\epsilon_0\in]0,a\alpha[$ and suppose that
$a,m,\alpha$ satisfy,
\begin{equation}\label{Hyp-m}
-\epsilon_0\leq\inf_{\xi\in\mathbb
R}\mu_1(a,m,\alpha;\xi)\leq\epsilon_0,
\end{equation}
then, if $m>1$, the set ${\rm M}(a,m,\alpha)$ is non-empty.
Moreover, if $\xi\in{\rm M}(a,m,\alpha)$, then,
\begin{equation}\label{|xi|<M}
|\xi|\leq
\sqrt{(1+a)\left(1-\frac1m\right)^{-1}\alpha},\end{equation} and the
eigenfunction $f_{\alpha,\xi}^{a,m}$ satisfies~:
$$\left(f_{\alpha,\xi}^{a,m}\right)'(0_{\pm})>0.$$
\end{thm}
\paragraph{\bf Proof.}
Notice that, if $m\leq1$, Formula (\ref{IIderVP-eq}) implies that
the function $\mapsto\mu_1(a,m,\alpha;\cdot)$ is strictly
decreasing. Therefore,
$$
\mu_1(a,m,\alpha;\xi)=\lim_{\xi\to+\infty}\mu^{(1)}(a,m,\alpha;\xi),$$
which proves (\ref{IICcr-eq}), thanks to (\ref{lim-pm-inf}).\\
Suppose now that $m>1$ and that Hypothesis (\ref{Hyp-m}) holds. We
denote by~:
\begin{equation}\label{d}
d=\sqrt{(1+a)\left(1-\frac1m\right)^{-1}\alpha}. \end{equation}
Notice that there exists $\xi_d\in[-d,d]$ such that,
$$\mu_1(a,m,\alpha;\xi_d)=\min_{\xi\in[-d,d]}\mu_1(a,m,\alpha;\xi).$$
As $m>1$, Formula (\ref{IIderVP-eq}) gives that the function
$\mu_1(a,m,\alpha;\cdot) $ is strictly increasing on the intervals
$]-\infty,-d]$ and $[d,+\infty[$. We then have,
$$\mu_1(a,m,\alpha;\xi_d)\leq\mu_1(a,m,\alpha,d)<
\mu_1(a,m,\alpha;\xi),\quad\forall\xi\in[d,+\infty],$$ and
$$
\mu_1(a,m,\alpha;\xi)>\frac1m+a\alpha>\epsilon_0,\quad\forall
\xi\in]-\infty,-d].
$$
This gives now,
$$
\mu_1(a,m,\alpha;\xi_d)=\inf_{\xi\in\mathbb R}\mu_1(a,m,\alpha;\xi),
$$
and hence the set ${\rm M}(a,m,\alpha)$ is non-empty and bounded.\\
Given $\xi\in{\rm M}(a,m,\alpha)$, the equation satisfied by
$f_{\alpha,\xi}^{a,m}$ in $\mathbb R_-$ can be written as~:
\begin{equation}\label{IIevEq-}\left\{\begin{array}{l}
-\left(f_{\alpha,\xi}^{a,m}\right)''+(t-\xi)^2f_{\alpha,\xi}^{a,m}=
m\left(\mu_1(a,m,\alpha;\xi)-a\alpha\right)
f_{\alpha,\xi}^{a,m},\quad t<0,\\
\left(f_{\alpha,\xi}^{a,m}\right)'(0_-)=\gamma(a,m,\alpha;\xi)f_{\alpha,\xi}^{a,m}(0),
\end{array}\right.\end{equation}
with $\mu_1(a,m,\alpha;\xi)-a\alpha<\epsilon_0-a\alpha<0$. Therefore
$\gamma(a,m,\alpha;\xi)>0$ and consequently
$\left(f_{\alpha,\xi}^{a,m}\right)'(0_\pm)>0$.\hfill$\Box$\\

We collect some useful relations in the next proposition.

\begin{prop}\label{II-M1M3}
Let $\eta\in{\rm M}(a,m,\alpha)$. Then we have,
\begin{equation}\label{II-M1}
\int_{\mathbb
R_+}(t-\eta)\left|f_{\alpha,\eta}^{a,m}(t)\right|^2dt+\frac1m
\int_{\mathbb
R_-}(t-\eta)\left|f_{\alpha,\eta}^{a,m}(t)\right|^2dt=0,
\end{equation}
\begin{eqnarray}\label{II-M3}
&&\int_{\mathbb
R_+}(t-\eta)^3\left|f_{\alpha,\eta}^{a,m}(t)\right|^2dt+\frac1m
\int_{\mathbb
R_-}(t-\eta)^3\left|f_{\alpha,\eta}^{a,m}(t)\right|^2dt\\
&&\quad=\frac16\left(1-\frac1m\right)\left|f_{\alpha,\eta}^{a,m}(0)\right|^2+
2\eta^2\left(\eta^2\left(1-\frac1m\right)-(a+1)\alpha\right)\left|f_{\alpha,\eta}^{a,m}(0)\right|^2
\nonumber\\
&&\quad\hskip0.4cm+\frac13\left(a-\frac1m\right)\alpha\int_{\mathbb
R_-}\left|f_{\alpha,\eta}^{a,m}(t)\right|^2dt.\nonumber
\end{eqnarray}
\end{prop}
\paragraph{\bf Proof.}
We denote by~:
$$L_+=-\partial_t^2+(t-\eta)^2-\alpha,\quad
L_-=\frac1m\left(-\partial_t^2+(t-\eta)^2\right)+a\alpha.$$ Let $p$
a polynomial, $f=f_{\alpha,\eta}^{a,m}$  and $v=2pf'-p'f$. By direct
calculation, we have~:
\begin{eqnarray*}
&&L_+v=\left(p^{(3)}-4p'\left[(t-\eta)^2-\alpha\right]-4p(t-\eta)\right)f,\\
&&L_-v=\frac1m\left(p^{(3)}-4p'\left[(t-\eta)^2+ma\alpha\right]-4(t-\eta)\right)f.
\end{eqnarray*}
Integrating by parts, we get,
\begin{eqnarray}\label{II-Ipp}
&&\int_{\mathbb R_+}f(t)\cdot(L_+f)(t)\,dt+\frac1m\int_{\mathbb R_-}
f(t)\cdot(L_-f)(t)\,dt\nonumber\\
&&\quad=\left(v'(0_+)-v'(0_-)\right)f(0)-\left(v(0_+)-v(0_-)\right)f'(0_+).
\end{eqnarray}
Taking $p=1$, we obtain (\ref{II-M1}). To obtain (\ref{II-M3}), we
take $p=(t-\eta)^2$. In this case,
\begin{eqnarray*}
&&\int_{\mathbb R_+}f(t)\cdot(L_+f)(t)\,dt+\frac1m\int_{\mathbb R_-}
f(t)\cdot(L_-f)(t)\,dt\nonumber\\
&&\quad=-2\left(1-\frac1m\right)|f(0)|^2+2\eta^2\left(f''(0_+)-\frac1mf''(0_-)\right)f(0),
\end{eqnarray*}
which is enough to deduce (\ref{II-M3}).\hfill$\Box$\\

\subsection{The function $\alpha(a,m)$}
In the theorem below, we define the function $\alpha(\cdot,\cdot)$
appearing in the statement of Theorem~\ref{k2Th1}.

\begin{thm}\label{IIestalp-rem}
Given $a,m>0$, there exists a unique solution $\alpha(a,m)$ of the
equation~:
\begin{equation}\label{IIestalp-rem'}
\inf_{\xi\in\mathbb R}\mu_1\left(a,m,\alpha(a,m);\xi\right)=0.
\end{equation}
Moreover, $\alpha(a,m)=1$ if $m\leq 1$, and $\Theta_0<\alpha(a,m)<1$
if $m>1$.
\end{thm}
\paragraph{\bf Proof.}
We start by proving the existence of $\alpha(a,m)$. The min-max
principle gives that the function
$$\alpha\mapsto\inf_{\xi\in\mathbb R}\mu_1(a,m,\alpha;\xi)$$
is Lipschitz. Lemma~\ref{IIestalp} gives immediately that
\begin{eqnarray*}
&&\inf_{\xi\in\mathbb R}\mu_1(a,m,\alpha;\xi)>0,\quad\forall
\alpha<\Theta_0,\\
&&\inf_{\xi\in\mathbb R}\mu_1(a,m,\alpha;\xi)<0,\quad\forall
\alpha>1.\end{eqnarray*} Therefore, by the intermediate value
theorem, there exists at least one solution
$\alpha=\alpha(a,m)\in[\Theta_0,1]$ satisfying
(\ref{IIestalp-rem'}).\\
If $m\leq1$, then by (\ref{IICcr-eq}), $\alpha(a,m)=1$ and hence it
is unique. If $m>1$, the function $\mu_1(a,m,\alpha;\cdot)$ is
increasing in $[d,+\infty[$ (cf. (\ref{IIderVP-eq}) and (\ref{d}))
and thus, thanks to (\ref{lim-pm-inf}), we have~:
$$\mu_1(a,m,\alpha;\xi)<1-\alpha,\quad \forall \xi\in[d,+\infty[,\quad \forall \alpha>0.$$
Consequently any solution of (\ref{IIestalp-rem'}) satisfies
$$\alpha(a,m)<1.$$
That $\alpha(a,m)>\Theta_0$ follows from the fact that ${\rm
M}(a,m,\alpha)$ is non-empty (cf. Theorem~\ref{min-thm1}). Actually,
let $\xi\in {\rm M}(a,m,\alpha)$ and let us look at the equation
satisfied by the eigenfunction $f_{\alpha,\xi}^{a,m}$ in $\mathbb
R_+$~:
\begin{equation}\label{IIL-ThG-eq}
\left\{\begin{array}{l}
-\partial_t^2f_{\alpha,\xi}^{a,m}+(t-\xi)^2f_{\alpha,\xi}^{a,m}=\alpha(a,m)
f_{\alpha,\xi}^{a,m},\quad t>0,\\
\left(f_{\alpha,\xi}^{a,m}\right)'(0_+)=\gamma(a,m,\alpha;\xi)f_{\alpha,\xi}^{a,m}(0).
\end{array}\right.
\end{equation}
Then,
$$
\alpha(a,m)\geq\lambda_1(\gamma(a,m,\alpha;\xi),\xi),
$$
where $\lambda_1(\gamma(a,m,\alpha;\xi),\xi)$ is defined by
(\ref{l1-ga,xi}). Theorem~\ref{min-thm1} gives
$\gamma(a,m,\alpha;\xi)>0$, and therefore $\alpha(a,m)>\Theta_0$.\\
It remains now to prove the uniqueness of $\alpha(a,m)$. Actually,
we need only to prove uniqueness when $m>1$. It is sufficient to
prove the following claim,
\begin{equation}\label{claim}
\text{ If }\inf_{\xi\in\mathbb R}\mu_1(a,m,\alpha;\xi)=0,\text{ then
}\forall \beta>\alpha,\quad \inf_{\xi\in\mathbb
R}\mu_1(a,m,\beta;\xi)<0.
\end{equation}
Notice that, after comparing the quadratic forms $Q[a,m,\alpha;\xi]$
and $Q[a,m,\beta;\xi]$, we get for any $\eta\in{\rm M}(a,m,\alpha)$,
\begin{equation}\label{var-alp-bet1}
Q[a,m,\beta;\eta](f_{\alpha,\eta}^{a,m})+(\beta-\alpha)\left(\int_{\mathbb
R_+}|f_{\alpha,\eta}^{a,m}(t)|^2dt-a\int_{\mathbb
R_-}|f_{\alpha,\eta}^{a,m}(t)|^2dt\right)=0.
\end{equation}
We have the following immediate consequence of the min-max
principle,
\begin{eqnarray*}
&&\left(\lambda^N_1(\eta)-\alpha\right)\int_{\mathbb
R_+}|f_{\alpha,\eta}^{a,m}(t)|^2dt+\left(\frac1m\lambda^N_1(-\eta)+a\alpha\right)\int_{\mathbb
R_-}|f_{\alpha,\eta}^{a,m}(t)|^2dt\\
&&\leq \mu_1(a,m,\alpha;\eta)\int_{\mathbb
R}|f_{\alpha,\eta}^{a,m}(t)|^2dt.\end{eqnarray*} Since
$\mu_1(a,m,\alpha;\eta)=0$ and $m>1$, the above estimate reads as
$$\frac{\Theta_0}m\leq\left(\int_{\mathbb
R_+}|f_{\alpha,\eta}^{a,m}(t)|^2dt-a\int_{\mathbb
R_-}|f_{\alpha,\eta}^{a,m}(t)|^2dt\right).$$ We have actually used
the fact that $\Theta_0<\alpha<1$ and that
$\lambda_1^N(\pm\eta)\geq\Theta_0$.\\
Therefore, thanks to (\ref{var-alp-bet1}), we get finally,
$$\mu_1(a,m,\beta;\eta)<0,$$
which proves Claim (\ref{claim}).
\hfill$\Box$\\

In the next proposition, we establish the monotonicity of
$\alpha(a,\cdot)$.
\begin{prop}\label{mont-alp}
The function $[1,+\infty[\ni m\mapsto\alpha(a,m)$ is strictly
decreasing.
\end{prop}
\paragraph{\bf Proof.}
As we are interested in the dependence on $m$, we omit $a$ from the
notation and we write $\alpha(m)$ for $\alpha(a,m)$. Suppose that~:
$$m_1>m_2>1.$$ Notice that,
$\forall\xi\in\mathbb R$, $\forall u\in B^1(\mathbb R)$, we have~:
\begin{eqnarray*}
Q[a,m_1,\alpha(m_1);\xi](u)&=&Q[a,m_2,\alpha(m_2);\xi](u)\\
&&+\left(\frac1{m_1}-\frac1{m_2}\right)\int_{\mathbb R_+}
\left\{|u'(t)|^2+|(t-\xi)u(t)|^2\right\}dt\\
&&+\left(\alpha(m_2)-\alpha(m_1)\right)\left( \int_{\mathbb
R_+}|u(t)|^2dt-a\int_{\mathbb R_-}|u(t)|^2dt\right).
\end{eqnarray*}
In particular, for $\xi=\eta\in{\rm M}(a,m_2,\alpha)$ and
$u=f_{\alpha(m_2)}$, we have~:
\begin{eqnarray*}
&&Q[a,m,\alpha(m_1);\eta](f_{\alpha(m_2)})\leq
\Theta_0\left(\frac1{m_1}-\frac1{m_2}\right)\int_{\mathbb
R_+}|f_{\alpha(m_2)}(t)|^2dt\\
&&+\left(\alpha(m_2)-\alpha(m_1)\right)\left(\int_{\mathbb
R_+}|f_{\alpha(m_2)}(t)|^2dt-a\int_{\mathbb
R_-}|f_{\alpha(m_2)}(t)|^2dt\right),
\end{eqnarray*}
where $f_{\alpha(m_2)}=f_{\alpha(m_2),\eta}^{a,m_2}$.\\ Suppose by
contradiction that $\alpha(m_1)\geq\alpha(m_2)$, we obtain, thanks
to (\ref{var-alp-bet1}) and the fact that $\alpha(m_2)<1$,
\begin{eqnarray*}
Q[a,m,\alpha(m_1);\eta](f_{\alpha(m_2)})&\leq&
\Theta_0\left(\frac1{m_1}-\frac1{m_2}\right)\int_{\mathbb
R_+}|f_{\alpha(m_2)}(t)|^2dt\\
&&+\left(\alpha(m_2)-\alpha(m_1)\right)\frac{\Theta_0}m.
\end{eqnarray*}
The min-max principle now gives~:
$$\inf_{\xi\in\mathbb R}\mu_1(a,m_1,\alpha(m_1);\xi)<0,$$
which is the desired contradiction. \hfill$\Box$\\

\subsection{Asymptotic analysis with respect to $m$}\label{subSec-m->infty} \ \\
We finish in this subsection the proof of Theorem~\ref{k2Th2}. We
start with the following technical lemma.

\begin{lem}\label{Unif-exp-dec}
Given $a,m>0$, let $\alpha=\alpha(a,m)$. For any
$\epsilon\in]0,1[$, there exists a constant $C>0$, such that,
$$\forall m>1,\quad\forall a>0,\quad \forall \xi\in{\rm
M}(a,m,\alpha),$$ the eigenfunction $f_{\alpha,\xi}^{a,m}$ decays in
the following way,
\begin{equation}\label{Unif-dec-eq}
\left\|\exp\left(\frac{\epsilon(t-\xi)^2}2\right)f_{\alpha,\xi}^{a,m}\right\|_{H^1(\mathbb
R_+)}+\frac1{\sqrt{m}}\left\|\exp\left(\frac{\epsilon(t-\xi)^2}2\right)f_{\alpha,\xi}^{a,m}\right\|_{H^1(\mathbb
R_-)}\leq C.
\end{equation}
\end{lem}
\paragraph{\bf Proof.}
Let us consider a function $\Phi\in H^1(\mathbb R)$. For simplicity
we shall write $f$ for $f_{\alpha,\xi}^{a,m}$. Given an integer
$N\in\mathbb N$, an integration by parts gives the following
identity~:
\begin{eqnarray}\label{II-1Agmeqn1-Th}
&&\\
&&\int_0^N \left[\left|\left(e^\Phi f\right)'\right|^2+
\left|(t-\xi)e^\Phi f\right|^2-\alpha\left|e^\Phi
f\right|^2\right]dt
-f'(N)e^{2\Phi(N)}f(N)\nonumber\\
&&\int_{-N}^0\left[\frac1m\left(\left|\left(e^\Phi
f\right)'\right|^2+ \left|(t-\xi)e^\Phi
f\right|^2\right)+a\alpha\left|e^\Phi
f\right|^2\right]dt+f'(-N)e^{\Phi(-N)}f(-N)\nonumber\\
&&=\mu_1(a,m,\alpha;\xi)\left\|e^\Phi f\right\|^2_{L^2([-N,N])}
+\left\|\Phi'e^\Phi
f\right\|^2_{L^2([0,N])}+\frac1m\left\|\Phi'e^\Phi
f\right\|^2_{L^2([-N,0])}.\nonumber
\end{eqnarray}
Let us recall that the eigenfunction $f$ is strictly positive and
that $\alpha$ and $\xi$ are selected in such a way that
$\mu_1(a,m,\alpha;\xi)=0$. It results then from the eigenvalue
equation satisfied by $f$~:
$$f''(t)>0,\quad
\forall t\in ]-\infty,\xi[\,\cup\,]\xi+\sqrt{\alpha},+\infty[.$$
Therefore, the function $f'$ is increasing on
$]-\infty,\xi[\,\cup\,]\xi+\sqrt{\alpha},+\infty[$. On the other hand,
as $f_{|_{\mathbb R_\pm}}\in H^2(\mathbb R_\pm)$, the Sobolev
Imbedding Theorem gives $\displaystyle\lim_{t\to+\infty}f'(t)=0$.
Thus, combining with the monotonicity of $f'$, we get,
$$f'(\pm N)<0,\quad \forall N>
\max(\xi+\sqrt{\alpha},-\xi).$$ It results now from
(\ref{II-1Agmeqn1-Th}) the following estimate~:
\begin{eqnarray}\label{II-1Agmeqn2-Th}
&&\int_0^N \left[\left|\left(e^\Phi f\right)'\right|^2+
\left((t-\xi)^2-\alpha-|\Phi'|\right)\left|e^\Phi f\right|^2\right]dt\\
&&+ \int_{-N}^0\left[\left(\left|\left(e^\Phi f\right)'\right|^2+
\left((t-\xi)^2-|\Phi'|^2\right)e^\Phi f\right)+a\alpha\left|e^\Phi
f\right|^2\right]dt\leq0.\nonumber
\end{eqnarray}
Now we take $\Phi$ as~:
$$\Phi(t)=\epsilon\frac{(t-\xi)^2}2.$$
We can then rewrite (\ref{II-1Agmeqn2-Th}) as~:
\begin{eqnarray}\label{II-1Agmeqn4-Th}
&&\int_{t\in[0,N],(t-\xi)\geq a_\epsilon}\left[\left|\left(e^\Phi
f\right)'\right|^2 +\left|e^\Phi
f\right|^2\right]dt\\
&&+\frac1m\int_{t\in[-N,0],(t-\xi)\geq a_\varepsilon}\left[\left|\left(e^\Phi
f\right)'\right|^2 +\left|e^\Phi f\right|^2\right]dt\leq e^{\epsilon
a_\epsilon/2},\nonumber
\end{eqnarray}
where $a_\epsilon>0$ satisfies~:
$$(1-\varepsilon^2)a_\epsilon^2-1\geq1.$$
Let us now take $C=e^{\epsilon a_\epsilon/2}$. Noticing that the
estimate (\ref{II-1Agmeqn4-Th}) is uniform with respect to $N$, we
get the result of the lemma by passing to the limit $N\to+\infty$.\hfill$\Box$\\

 In the next proposition we
give a `rough' asymptotic result.

\begin{prop}\label{lim1+,+infty}
Given $a>0$, the following asymptotics hold~:
$$\lim_{m\to1^+}\alpha(a,m)=1\quad\text{ and }\quad\lim_{m\to+\infty}\alpha(a,m)=\Theta_0.$$
Moreover, there exists a function $\epsilon(m)$ such that,
\begin{equation}\label{Br10''}
\epsilon(m)>0,\quad\lim_{m\to+\infty}\epsilon(m)=0\text{ \rm and }{\rm M}\left(a,m,\alpha(a,m)\right)\subset
]-\epsilon(m)+\xi_0,\xi_0+\epsilon(m)[.
\end{equation}
\end{prop}
\paragraph{\bf Proof.} In the proof of this proposition and in the
sequel of this section, we shall write $\alpha$ for $\alpha(a,m)$.\\
{\bf The limit $m\to1^+$.}\\
Let $\epsilon\in]0,1[$ and $m=1+\epsilon$. Noticing that
$\frac1m>1-\epsilon$, we get for any function $u\in B^1(\mathbb R)$,
\begin{eqnarray}\label{Q-m1}
Q[a,m,\alpha;\xi](u)&\geq&\int_{\mathbb
R}\left(|u'(t)|^2+|(t-\xi)u(t)|^2-\alpha|u(t)|^2\right)dt\\
&&-\epsilon\int_{\mathbb
R_-}\left(|u'(t)|^2+|(t-\xi)u(t)|^2\right)dt.\nonumber\\
\end{eqnarray}
Using the fact (cf. (\ref{Th-0}))
$$\int_{\mathbb
R_+}\left(|u'(t)|^2+|(t-\xi)u(t)|^2\right)dt\geq\Theta_0\int_{\mathbb
R_+}|u(t)|^2dt,$$ together with $\alpha>\Theta_0$, we rewrite
(\ref{Q-m1}) as,
\begin{eqnarray*}
(1+m\epsilon)Q[a,m,\alpha;\xi](u)&\geq&\int_{\mathbb
R}\left[|u'(t)|^2+|(t-\xi)u|^2\right]dt-\alpha\int_{\mathbb
R}|u(t)|^2dt\\
&&+m\epsilon(\Theta_0-\alpha)\int_{\mathbb R}|u(t)|^2dt.
\end{eqnarray*}
Applying the min-max principle, we obtain,
$$(1+m\epsilon)\mu_1(a,m,\alpha;\xi)\geq 1-\alpha+m\epsilon(\Theta_0-\alpha).$$
Taking the infimum with respect to $\xi$, we obtain, thanks to the
definition of $\alpha$ in Theorem~\ref{IIestalp-rem},
$$0\geq 1-\alpha+m(\Theta_0-\alpha)\epsilon.$$
Recalling that $\alpha<1$, this is sufficient to deduce the required
limit.\\
{\bf The limit $m\to+\infty$.}\\
Recall that $\varphi_0$ is the first eigenfunction of $\mathcal
L^N[\xi_0]$ (cf. Subsection~\ref{ODE}). Let $\widetilde\varphi_0$ be
the even extension of $\varphi_0$ in $\mathbb R$, i.e.
$$\widetilde\varphi_0(t)=\left\{\begin{array}{l}
\varphi_{0}(t),\quad
t>0\\
\varphi_0(-t),\quad t<0.\end{array}\right.$$ Let $\chi$ be a cut-off
function such that
\begin{equation}\label{chi-1}
0\leq\chi\leq1,\quad\chi=0\quad \text{in }]-\infty,-1],\quad
\text{and }\chi=1\quad \text{in }[0,+\infty[.
\end{equation}
There exist constants $C,m_0>0$ such that,
\begin{equation}\label{II-chi}
Q[a,m,\alpha;\xi_0]\left(\chi(\sqrt{m}\,
t)\,\widetilde\varphi_0\right)\leq
\Theta_0-\alpha+\frac{C}{\sqrt{m}},\quad\forall m\geq m_0.
\end{equation}
It results now from the min-max principle and our choice of $\alpha$
that
$$Q[a,m,\alpha;\xi_0]\left(\chi(\sqrt{m}\, t)\,\widetilde\varphi_0\right)\geq0.$$
Therefore, (\ref{II-chi}) reads,
\begin{equation}\label{ii-mu-1}
0\leq \Theta_0-\alpha+\frac{C}{\sqrt{m}},\quad\forall m\geq m_0.
\end{equation}
Remembering that $\alpha>\Theta_0$, we get finally that,
$$\Theta_0< \alpha\leq\Theta_0+\frac{C}{\sqrt{m}},$$
and consequently,
$\displaystyle\lim_{m\to+\infty}\alpha(a,m)=\Theta_0$.\\
{\bf Localization of ${\rm M}(a,m,\alpha)$.}\\
Let $\xi\in {\rm M}(a,m,\alpha(a,m))$. We denote by
$\gamma_\xi=\gamma(a,m,\alpha(a,m);\xi)$. The equation
$\partial_\xi\mu_1(a,m,\alpha;\xi)=0$ gives the following relation~:
\begin{equation}\label{IIPdG}
\gamma_\xi^2=\frac{a+1}{m-1}\alpha-\frac1m\xi^2.
\end{equation}
Therefore, thanks to Theorem~\ref{min-thm1}, there exist  constants
$C,m_0$  such that,
\begin{equation}\label{gam-m-bd}
0<\gamma_\xi\leq \frac{C}{\sqrt{m}},\quad\forall m\geq m_0.
\end{equation}
We define the function~:
$$\phi_\xi=\frac1{\|f_{\alpha,\xi}^{a,m}\|_{L^2(\mathbb R_+)}}
e^{-\gamma_\xi t}f_{\alpha,\xi}^{a,m}(t).$$ Using
Lemma~\ref{Unif-exp-dec}, we get,
\begin{equation}\label{N(phi-xi)=1}
\forall\delta\in]0,1/2[,\quad \exists\,C_\delta,m_\delta>0\text{
s.t. }\forall m\geq m_\delta,\quad1-C_\delta
m^{\delta-1/2}\leq\|\phi_\xi\|_{L^2(\mathbb R_+)}\leq1.
\end{equation}
It results also from the eigenvalue equation satisfied by
$f_{\alpha,\xi}^{a,m}$,
\begin{equation}\label{ThII-Meq}
\left\{
\begin{array}{l}
-\phi''_\xi(t)+(t-\xi)^2\phi_\xi(t)=\alpha(a,m)\phi_\xi(t)+\gamma_\xi^2\phi_\xi(t)
+2\gamma_\xi\phi_\xi'(t),\quad
t>0,\\
\phi_\xi'(0)=0. \end{array} \right. \end{equation} This yields the
following estimate (cf. (\ref{k2-qfk1})),
$$q[0,\xi](\phi_\xi)\leq \alpha(a,m)+C\gamma_\xi.$$
Using the min-max principle, (\ref{ii-mu-1}), (\ref{gam-m-bd}), we
get the following upper bound~:
$$\lambda_1^N(\xi)\leq\Theta_0+\frac{C}{m^{\delta}},\quad
\forall m\in[m_\delta,+\infty[.$$ Remembering the definition of
$\Theta_0$ in (\ref{Th-gam}) and (\ref{Th-0}), we get
$\lambda_1^N(\xi)\geq\Theta_0$. Therefore, we have, after applying
Taylor's Formula to $\lambda_1^N(\cdot)$ up to the order $2$ near
$\xi_0$,
$$|\xi-\xi_0|\leq \frac{C}{m^{\delta/2}}.$$
This achieves the proof of the proposition. Let us mention also that
it results now from the relation (\ref{IIPdG}),
\begin{equation}\label{gam-m}
\gamma_\xi=\frac{\sqrt{a\Theta_0}}{\sqrt{m}}(1+o(1)),\quad
(m\to+\infty). \end{equation}
\hfill$\Box$\\

The following lemma is very useful for the localization of the set
${\rm M}(a,m,\alpha)$ when $m\to+\infty$.

\begin{lem}\label{II-LdeGe-Gen}
Given $a>0$, there  exist $m_0>1$ and a function
$m\mapsto\epsilon(m)$ satisfying
$$\epsilon(m)>0,\quad\lim_{m\to+\infty}\epsilon(m)=0,$$
such that,  if $m>m_0$, we have~:
\begin{equation}\label{gam>0}
\gamma(a,m,\alpha;\xi)>0,\quad\forall\xi\in]-\epsilon(m)+\xi_0,\xi_0+\epsilon(m)[,
\end{equation}
and
\begin{equation}\label{EqDe-alpha}
\alpha+\mu_1(a,m,\alpha;\xi)=\lambda_1\left(\gamma(a,m,\alpha;\xi),\xi\right),
\quad\forall \xi\in]-\epsilon(m)+\xi_0,\xi_0+\epsilon(m)[.
\end{equation}
\end{lem}
\paragraph{\bf Proof.}
Let $\xi\in\mathbb R_+$. Looking at the eigenvalue equation
satisfied by $f_{\alpha,\xi}^{a,m}$, we get~:
$$\left\{\begin{array}{l}
-\left(f_{\alpha,\xi}^{a,m}\right)''(t)+(t-\xi)^2f_{\alpha,\xi}^{a,m}(t)=m(
\mu_1(a,m,\alpha;\xi)-a\alpha)f_{\alpha,\xi}^{a,m},\quad t<0,\\
\left(f_{\alpha,\xi}^{a,m}\right)'(0_-)=m\,\gamma(a,m,\alpha;\xi)f_{\alpha,\xi}^{a,m}(0).
\end{array}\right.$$
When $|\xi-\xi_0|<\epsilon(m)$, it results from the min-max
principle the existence of $m_0>0$ such that~:
$$\forall m\geq m_0,\quad \mu_1(a,m,\alpha;\xi)\leq \frac12a\Theta_0.$$
To obtain the above estimate, it is enough to use the function
$\chi(\sqrt{m}t)\widetilde\varphi_\xi(t)$ as a quasi-mode
(cf.~(\ref{chi-1})), where the function
$\varphi_\xi=\varphi_{0,\xi}$ is the eigenfunction associated with
$\mu^N(\xi)$ (cf. (\ref{l-ga=0})), and
$\widetilde\varphi_\xi$ is the even extension of $\varphi_\xi$.\\
Therefore, remembering that $\alpha>\Theta_0$,
$$\mu_1(a,m,\alpha;\xi)-a\alpha<0\quad \forall m\geq m_0,$$
and consequently we obtain (\ref{gam>0}). Looking now at the
eigenvalue equation in $\mathbb R_+$~:
$$\left\{\begin{array}{l}
-\left(f_{\alpha,\xi}^{a,m}\right)''(t)+(t-\xi)^2f_{\alpha,\xi}^{a,m}(t)=
(\mu_1(a,m,\alpha;\xi)+\alpha)f_{\alpha,\xi}^{a,m},\quad t>0,\\
\left(f_{\alpha,\xi}^{a,m}\right)'(0_+)=\gamma(a,m,\alpha;\xi)f_{\alpha,\xi}^{a,m}(0),
\end{array}\right.$$
with, thanks to (\ref{IIestalp-eq}),
$$\alpha+\mu_1(a,m,\alpha;\xi)\leq\lambda^D_1(\xi).$$ As
$\gamma(a,m,\alpha;\xi)>0$ when $m\geq m_0$, then, thanks to
Lemma~\ref{HeRu}, we obtain formula (\ref{EqDe-alpha}).\hfill$\Box$\\

In the next lemma we give a two-term asymptotics to $\alpha(a,m)$.
\begin{lem}\label{asy-alp}
The following asymptotic expansion holds,
\begin{equation}\label{asy-alp-eq}
\alpha(a,m)=\Theta_0+3C_1\frac{\sqrt{a\Theta_0}}{\sqrt{m}}+\mathcal
O\left(\frac{1}{m}\right),\quad (m\to+\infty),
\end{equation}
where the constant $C_1$ is defined in (\ref{C1}).
\end{lem}

\begin{rem}
We believe that $\alpha(a,m)$ has a complete asymptotic expansion in
powers of $\frac1{\sqrt{m}}$ as $m\to+\infty$.
\end{rem}
\paragraph{\bf Proof of Lemma~\ref{asy-alp}.}
Let $\xi\in{\rm M}(a,m,\alpha)$ and $\gamma=\gamma(a,m,\alpha;\xi)$.
It is sufficient to establish, thanks to (\ref{Th'-gam}), (\ref{C1})
and (\ref{gam-m}), the existence of constants $m_0,C>0$ such that,
\begin{equation}\label{II-al-Th}
\left|\alpha-\Theta(\gamma)\right|\leq \frac{C}{m},\quad \forall
m\geq m_0.
\end{equation}
Let us recall that $\mu_1(a,m,\alpha;\xi)=0$. The definition of
$\Theta(\gamma)$ (cf. (\ref{Th-gam})) together with
(\ref{EqDe-alpha}) gives the following lower bound for $\alpha$~:
$$\Theta(\gamma)\leq \alpha.$$
We look now for an upper bound. Consider the following quasi-mode,
$$
u(t)=\left\{
\begin{array}{l}
\varphi_{\gamma}(t),\quad t>0,\\
\varphi_{\gamma}(0)\exp(b_mt),\quad t<0,
\end{array}\right.
$$
where $\varphi_\gamma$ is defined in (\ref{varphi-gam}) and the parameter $b_m>0$
is to be chosen appropriately.\\
Let us notice that $u$ is in the form domain of $Q[a,m,\alpha;\xi]$,
and that,
\begin{eqnarray*}
&&\int_{-\infty}^0\left|\left(e^{b_mt}\right)'\right|^2dt=\frac{b_m}2,\quad
\int_{-\infty}^0e^{2b_mt}dt=\frac{1}{2b_m},\\
&&\int_{-\infty}^0(t-\xi(\gamma))^2e^{2b_mt}dt=\frac1{4b_m^2}
+\frac{\xi(\gamma)(1+\xi(\gamma))}{2b_m}.
\end{eqnarray*}
Hence, we obtain,
\begin{eqnarray}\label{II-j4}
Q[a,m,\alpha;\xi(\gamma)](u)&\leq& \Theta(\gamma)-\alpha
+\left(\frac{b_m}{2m}-\gamma+\frac{a\alpha}{2b_m}\right)|\varphi_{\gamma}(0)|^2\\
&&+\frac{1}m\left(\frac1{4b_m^2}+\frac{\xi(\gamma)(1+\xi(\gamma))}{2b_m}\right)
|\varphi_{\gamma}(0)|^2.\nonumber
\end{eqnarray}
On the other hand, thanks to the min-max principle and the choice of
$\alpha$, we have $$Q[a,m,\alpha;\xi(\gamma)](u)\geq0.$$
Let us choose $b_m$ in the form~:
$$b_m=b_0\sqrt{m},\quad \text{ with }b_0\geq0.$$
Noticing that $|\varphi_\gamma(0)|$ is bounded, thanks to
(\ref{gam-m-bd}), Formula (\ref{II-j4}) can be rewritten as~:
$$\alpha\leq\Theta(\gamma)
+\left(\frac{b_m}{2m}-\gamma+\frac{a\alpha}{2b_m}\right)|\varphi_{\gamma}(0)|^2
+\frac{C_0}{m}.$$ Having in mind (\ref{gam-m}), we obtain,
$$\left|\left(
\frac{b_m}{2m}-\gamma+\frac{a\alpha}{2b_m}\right)-
\frac{(b_0-(a\Theta_0)^{1/2})^2}{2b_0\sqrt{m}}\right|\leq
\frac{C}{m}.$$ Optimizing over $b_0$ leads to the choice
$b_0=(a\Theta_0)^{1/2}$. We get therefore the following upper bound,
$$\alpha\leq\Theta(\gamma)+\frac{C}{m},$$
and thus, we achieved the proof of the lemma.\hfill$\Box$\\

We give now a fine localization of the set ${\rm M}(a,m,\alpha)$.

\begin{prop}\label{lim+infty}
Given  $a>0$, there exist constants $C,m_0>0$ such that,
\begin{equation}\label{II-loc-Min}
\forall m\geq m_0,\quad{\rm M}(a,m,\alpha)\subset
\left]\xi_0+\frac{b}{\sqrt{m}}-\frac{C}{m},\,
\xi_0+\frac{b}{\sqrt{m}}+\frac{C}{m}\right[,
\end{equation}
where the constant $b>0$ is defined by,
\begin{equation}\label{II-def-b}
b=\sqrt{a}\frac{3C_1(1-3C_1)}{2(2-3C_1)},
\end{equation}
and the constant $C_1>0$ is introduced in (\ref{C1}).
\end{prop}
\paragraph{\bf Proof.}
Let us take $\xi\in {\rm M}(a,m,\alpha)$. It is sufficient to prove
the existence of constants $C>0$, $m_0>0$, independent of  $m$ and
such that,
\begin{equation}\label{II-re-so}
\left|\xi-\frac{b}{\sqrt{m}}\right|\leq \frac{C}m.
\end{equation}
Let $\gamma=\gamma(a,m,\alpha;\xi)$. Using (\ref{EqDe-alpha}), we
get $\alpha=\lambda_1(\gamma,\xi)$. Upon applying Taylor's formula
up to the order $2$ to the  function $\lambda_1(\cdot,\cdot)$ near
$(\gamma,\xi(\gamma))$, we get, thanks also to
Proposition~\ref{lim1+,+infty},
\begin{equation}\label{Br7}
(\xi-\xi(\gamma))^2=\frac1{\xi_0}(\alpha-\Theta(\gamma))(1+o(1)),\quad
\text{as }m\to+\infty.
\end{equation}
This  gives now, thanks to (\ref{II-al-Th}),
$$|\xi-\xi(\gamma)|\leq \frac{C}{m}.$$
Consequently, Taylor's Formula applied to the function
$\lambda_1(\cdot,\cdot)$ at $(0,\xi_0)$ will give as $m\to+\infty$
the following asymptotics,
\begin{eqnarray}
\alpha&=&\lambda_1(\gamma,\xi)\label{Br8}\\
&=&\Theta_0+a_1\gamma+a_2'\gamma^2+b_1\gamma(\xi-\xi_0)
+c_1(\xi-\xi_0)^2\nonumber\\
&&+\mathcal O(\gamma^3+|\xi-\xi_0|^{3}),\nonumber
\end{eqnarray}
where the  coefficients $a_1,a_2',b_1,c_1$ are defined by,
\begin{eqnarray*}
&&a_1=\left(\partial_\gamma\lambda_1\right)(0,\xi_0)=\Theta'(0),\quad
a_2'=\frac12\left(\partial_\gamma^2\lambda_1\right)(0,\xi_0),\\
&&b_1=\frac12\left(\partial_\gamma\partial_\xi\lambda_1\right)
(0,\xi_0),\quad c_1=\xi_0.\nonumber
\end{eqnarray*}
One has also, thanks to Taylor's Formula (applied to the function
$\Theta(\gamma)$) and to (\ref{xi(gam)}) ,
\begin{eqnarray}
\Theta(\gamma)&=&\Theta_0+a_1\gamma+a_2\gamma^2+\mathcal O
(\gamma^3),\label{Br9}\\
\xi(\gamma)&=&\xi_0+\frac{a_1}{2\xi_0}\gamma+\mathcal
O(\gamma^2),\label{Br10}
\end{eqnarray}
where the coefficients $a_1,a_2$ are defined by,
\begin{eqnarray*}
&&a_1=\Theta'(0),\quad a_2=\frac12\Theta''(0).
\end{eqnarray*}
By writing
$(\xi-\xi(\gamma))^2=(\xi-\xi_0)^2+(\xi-\xi(\gamma))^2-2(\xi-\xi_0)(\xi-\xi(\gamma))$,
we obtain thanks to (\ref{Br10''}) and  (\ref{Br10}),
\begin{equation}\label{Br10'}
(\xi-\xi(\gamma))^2=(\xi-\xi_0)^2+(\xi(\gamma)-\xi_0)^2
-\frac{a_1}{\xi_0}\gamma(\xi-\xi_0)+ o(\gamma^2).
\end{equation}
Using the formulas of differentiation in (\ref{Th7.5.5}) and
(\ref{Th7.5.6}), we get,
\begin{eqnarray}\label{Th''}
&&b_1=-\frac12a_1^2,\nonumber\\
&&\Theta''(\gamma)=\left(\partial_\gamma^2\lambda_1\right)(\gamma,\xi(\gamma))
+\xi'(\gamma)\left(\partial_{\gamma}\partial_\xi\lambda_1\right)(\gamma,\xi(\gamma)),\\
&&a_2'=a_2+\frac{a_1^3}{4\xi_0}.\nonumber
\end{eqnarray}
Solving the equation (\ref{Br7}), we get finally,
$$\xi-\xi_0=\frac{a_1(1-a_1)}{4\xi_0\left(1-\frac{a_1}2\right)}\gamma(1+o(1)),\quad
(m\to+\infty).$$ Recalling (\ref{gam-m}), we obtain
(\ref{II-re-so}). Finally, it is proved  in~\cite[(2.13)]{FoHe} that $a_1<1$ (and hence $b>0$).
\hfill$\Box$\\

\begin{rem}\label{b>0}
The following numerical estimate was obtained  in~\cite[Formula
(2.125)]{BonTh}~:
$$0.858\leq3C_1\leq0.888.$$
This shows that $b>0$.
\end{rem}

The asymptotic behavior of $\alpha(a,m)$ as $m\to+\infty$ permits
one to prove the existence of a unique minimum of the function
$\mu_1(a,m,\alpha;\cdot)$ for large values of the parameter $m$.

\begin{thm}\label{Unic-xi}
Given  $a>0$, there exists  $m_0>1$ such that, for any $m\geq m_0$,
the function $\xi\mapsto\mu_1\left(a,m,\alpha(a,m);\cdot\right)$ has
a unique non-degenerate positive minimum denoted by $\xi(a,m)$.
\end{thm}
\paragraph{\bf Proof.}
Let us take, thanks to Theorems~\ref{min-thm1}~and~\ref{lim+infty},
a critical point
$$\xi\in\left]\xi_0+\frac{b}{\sqrt{m}}-\frac{C}{m},\,
\xi_0+\frac{b}{\sqrt{m}}+\frac{C}{m}\right[.$$ It is sufficient to
prove that,
\begin{equation}\label{II-dr2>0}
\partial_\xi^2\mu_1(a,m,\alpha;\xi)>0.
\end{equation}
Notice that, Formula (\ref{IIderVP-eq}) gives,
\begin{equation}\label{II-der-mu(xi)}
\partial_\xi^2\mu^{(1)}(a,m,\alpha;\xi)=2(m-1)
\left[\gamma'(a,m,\alpha;\xi)\gamma(a,m,\alpha;\xi)+\frac1m\xi\right]
|f_{\alpha,\xi}^{a,m}(0)|^2.\end{equation}
Differentiating both
sides of (\ref{EqDe-alpha}), we get, thanks to (\ref{Th7.5.5}),
\begin{equation}\label{II-der-gam(xi)}
\gamma'(a,m,\alpha;\xi)=|\gamma(a,m,\alpha;\xi)|^2-\xi^2+\alpha+\mu_1(a,m,\alpha;\xi).
\end{equation}
Notice that, thanks to (\ref{gam-m}) and (\ref{asy-alp-eq}),
$$|\gamma(a,m,\alpha;\xi)|^2-\xi^2+\alpha=\left(a_1-2\frac{b}{\sqrt{a}\,\xi_0}\right)\gamma(1+o(1)),\quad
(m\to+\infty),$$ with $a_1=3C_1$. Since, $0<a_1<1$
(cf.~\cite[(2.13)]{FoHe}), we get, thanks to (\ref{II-def-b}),
$$a_1-2\frac{b}{\sqrt{a}\,\xi_0}>0.$$
We have therefore achieved the proof of the theorem.\hfill$\Box$\\

Using the regularity result in Theorem~\ref{reg}, one gets
immediately the following consequence of
Theorems~\ref{min-thm1}~and~\ref{Unic-xi}.

\begin{thm}\label{IIRe-Gen-Th}
Given $a>0$, there exists $m_0>1$ such that~:
$$\forall m\geq m_0,\quad\exists\,\epsilon_0(m)>0,
$$
and if $\widehat\alpha$ satisfies~:
$$
|\widehat\alpha-\alpha(a,m)|\leq\epsilon_0(m),$$ then the function
$$\xi\mapsto\mu_1(a,m,\widehat\alpha;\xi)$$
has a unique positive non-degenerate minimum which we denote by
$\widehat\xi(a,m)$.
\end{thm}

\section{Analysis of a `refined' family of model
operators}\label{Sec-MF}

\subsection{Notation and main theorem}
For the analysis of `curvature effects' (Theorem~\ref{k2Th3}), we
need to introduce a refined family of model operators. Let us
consider
\begin{equation}
a,m,\widehat\alpha,h\in\mathbb R_+,\quad\beta,\xi\in\mathbb R\quad
{\rm and} \quad\delta\in]\frac14,\frac12[.
\end{equation}
We assume further the following condition on $\beta,h$ and $\delta$,
$$\beta h^\delta<1.$$
We consider the quadratic form~:
$$
H^1_0(]-h^{\delta-1/2},h^{\delta-1/2}[)\ni u\mapsto
q_{h,\beta,\xi}^{a,m,\widehat\alpha}(u),
$$
defined by~
\begin{eqnarray}\label{II-Fq-M}
&&q_{h,\beta,\xi}^{a,m,\widehat\alpha}(u)=\int_0^{h^{\delta-1/2}}{\big
[} |u'(t)|^2 +(1+2\beta h^{1/2}t) \left|\left(t-\xi-\beta h^{1/2}
\frac{t^2}{2}\right)u(t)\right|^2\\
&&\hskip3cm-\widehat\alpha|u(t)|^2{\big ]}(1-\beta h^{1/2}t)dt\nonumber\\
&&\hskip1.5cm+\frac1m\int_{-h^{\delta-1/2}}^0{\big [} |u'(t)|^2
+(1+2\beta h^{1/2}t) \left|\left(t-\xi-\beta h^{1/2}
\frac{t^2}{2}\right)u(t)\right|^2\nonumber\\
&&\hskip3cm+a\widehat\alpha|u(t)|^2{\big ]}(1-\beta
h^{1/2}t)dt.\nonumber
\end{eqnarray}
By Friedrichs' Theorem, we associate to the quadratic form
$q^{a,m,\widehat\alpha}_{h,\beta,\xi}$ a non-bounded self-adjoint
operator $\mathcal H_{h,\beta,\xi}^{a,m,\widehat\alpha}$ on the
space
$$L^2(]-h^{\delta-1/2},h^{\delta-1/2}[;(1-\beta
h^{1/2}t)dt).$$ The domain of $\mathcal
H_{h,\beta,\xi}^{a,m,\widehat\alpha}$ is defined by~:
\begin{eqnarray}\label{II-D-opM}
&&\\
D(\mathcal H_{h,\beta,\xi}^{a,m,\widehat\alpha})&=&\{u\in
H^1_0(]-h^{\delta-1/2},h^{\delta-1/2}[);\quad
u_{|_{]-h^{\delta-1/2},0[}}\in H^2(]-h^{\delta-1/2},0[),\nonumber\\
&&\hskip0.2cm u_{|_{]0,h^{\delta-1/2}[}}\in
H^2(]0,h^{\delta-1/2}[);\quad u'(0_+)=\frac1m u'(0_-)\}.\nonumber
\end{eqnarray}
For $u\in D(\mathcal H_{h,\beta,\xi}^{a,m,\widehat\alpha})$, we
have~:
\begin{equation}\label{II-opM}
\left(\mathcal
H_{h,\beta,\xi}^{a,m,\widehat\alpha}u\right)(t)=\left\{
\begin{array}{l}
\left(H_{h,\beta,\xi}u-\widehat\alpha u\right)(t),\quad {\rm if} \quad t<0,\\
\left(\frac1m H_{h,\beta,\xi}u+a\widehat\alpha u\right)(t),\quad
{\rm if}\quad t>0,
\end{array}
\right.
\end{equation}
where $H_{h,\beta,\xi}$ is the differential operator~:
\begin{eqnarray}\label{II-OpC}
H_{h,\beta,\xi}&=&-\partial_t^2+(t-\xi)^2\\
&&+\beta h^{1/2}(1-\beta h^{1/2}t)^{-1}\partial_t+2\beta
h^{1/2}t\left(t-\xi-\beta h^{1/2}\frac{t^2}2\right)^2\nonumber\\
&&-\beta h^{1/2}t^2(t-\xi)+\beta^2h\frac{t^4}4.\nonumber
\end{eqnarray}
We denote by $\mu_j(\mathcal H_{h,\beta,\xi}^{a,m,\widehat\alpha})$
the increasing sequence of eigenvalues of $\mathcal
H_{h,\beta,\xi}^{a,m,\widehat\alpha}$. We are interested in finding
a lower bound of
$$\inf_{\xi\in\mathbb R}\mu_1\left(\mathcal
H_{h,\beta,\xi}^{a,m,\widehat\alpha}\right).$$ We shall always work
under the following general hypothesis,
\begin{equation}
m\geq m_0,
\end{equation}
where the constant $m_0>1$ fills the assumption of
Theorem~\ref{IIRe-Gen-Th}. We write,
\begin{equation}\label{II-alp-am}
\alpha=\alpha(a,m).
\end{equation}
We suppose also that (cf. Theorem~\ref{IIRe-Gen-Th})~:
$$|\widehat\alpha-\alpha|\leq \epsilon_0(m).$$
We denote then by $\widehat\eta$ and $\eta$ the unique numbers
defined by Theorem~\ref{IIRe-Gen-Th} such that~:
\begin{eqnarray}
\widehat\eta\in{\rm M}(a,m,\widehat\alpha),\quad \eta\in{\rm
M}(a,m,\alpha),\label{II-eta}
\end{eqnarray}
and we recall that $\widehat\eta,\eta>0$. Finally, we denote by~:
\begin{equation}\label{II-f-eta-n}
\widehat f=f_{\widehat\alpha,\widehat\eta}^{a,m},\quad
f=f_{\alpha,\eta}^{a,m}
\end{equation}
the positive eigenfunctions (and normalized for the $L^2$-norm in
$\mathbb R$) associated to $\mu_1(a,m,\widehat\alpha;\widehat\eta)$
and $\mu_1(a,m,\alpha;\eta)$
respectively.\\
We define the following functions~:
\begin{equation}\label{II-C1a,m}
\widetilde{\mathcal C}_1(a,m)=\int_{\mathbb
R_+}(t-\eta)^3|f(t)|^2dt+\frac1m\int_{\mathbb
R_-}(t-\eta)^3|f(t)|^2dt
-\frac12\left(1-\frac1m\right)|f(0)|^2,
\end{equation}
\begin{equation}\label{II-b0a,m}
b_1(a,m)=\int_{\mathbb R_+}|f(t)|^2dt-a\int_{\mathbb R_-}|f(t)|^2dt.
\end{equation}
Notice that, thanks to (\ref{II-M3}) and the asymptotic behavior as
$m\to+\infty$, the constant $\widetilde{\mathcal C}_1(a,m)$ is
negative for large values
of the parameter $m$.\\
Our aim in this section is to prove the following theorem.
\begin{thm}\label{II-thmF}
Given $a>0$ and $m\geq m_0$, then for every $M>0$, there exist
constants $C,\epsilon,h_0>0$ such that~:
$$\forall\beta\in]-M,M[,\quad \forall\xi\in\mathbb R,\quad
\forall
\widehat\alpha\in[-\epsilon+\alpha,\alpha+\epsilon],\quad\forall
h\in]0,h_0],$$ we have~:
\begin{equation}\label{II-thmF-eq}
\mu_1(\mathcal H_{h,\beta,\xi}^{a,m,\widehat\alpha})\geq
b_1(a,m)(\widehat\alpha-\alpha)+\widetilde{\mathcal C}_1(a,m)\beta h^{1/2}
-C\left[|\widehat\alpha-\alpha|^2+h^{\delta+1/4}\right].
\end{equation}
\end{thm}

Let us mention that in all the estimates of this section, we do not
seek to control the constants uniformly with respect to the
parameter $m$.

\subsection{A first order approximation of
$\mu_1(a,m,\widehat\alpha;\widehat\eta)$} We construct a first order
approximation of $\mu_1(a,m,\widehat\alpha;\widehat\eta)$ by the
help of an approximate eigenfunction. We recall first the definition
of the `regularized resolvent'. Given $a,m,\beta>0$ and
$\xi\in\mathbb R$, the regularized resolvent $\mathcal
R[a,m,\beta;\xi]$ is the bounded linear operator on $L^2(\mathbb R)$
defined by,
\begin{equation}\label{reg-res}
\mathcal R[a,m,\beta;\xi]u=\left\{
\begin{array}{l}
\left(H[a,m,\beta;\xi]-\mu_1(a,m,\beta;\xi)\right)^{-1}u,\quad {\rm
if}\quad u\bot f_{\beta,\xi}^{a,m},\\
0,\quad{\rm if }\quad u\in\mathbb R\cdot f_{\beta,\xi}^{a,m}.
\end{array}\right.
\end{equation}

\begin{lem}\label{II-mu-th'}
There exist constants $C,\epsilon>0$ such that, if
$|\widehat\alpha-\alpha|\leq\epsilon$, we have~:
\begin{equation}\label{II-eq-mu-th'}
\left|
\mu_1(a,m,\widehat\alpha;\widehat\eta)-\mu_1(a,m,\alpha;\widehat\eta)
-\widehat b_1(a,m)(\widehat\alpha-\alpha)\right|\leq
C|\widehat\alpha-\alpha|^2,
\end{equation}
where
$$\widehat b_1(a,m)=\int_{\mathbb R_+}|f_{\alpha,\widehat\eta}^{a,m}(t)|^2dt-a\int_{\mathbb
R_-}|f_{\alpha,\widehat\eta}^{a,m}(t)|^2dt.$$
\end{lem}
\paragraph{\bf Preuve.}
Let $\tau=\widehat\alpha-\alpha$. We look for $b_0,b_1\in\mathbb R$
and $u_0,u_1\in L^2(\mathbb R)$ such that,
$$\left(H[a,m,\widehat\alpha;\widehat\eta]-(b_0+b_1\tau)\right)(u_0+\tau
u_1)\sim 0\quad \text{ in }\mathbb R.$$ Let us write,
\begin{eqnarray*}
&&\left(H[a,m,\widehat\alpha;\widehat\eta]-(b_0+b_1\tau)\right)(u_0+\tau
u_1)\\
&&=\left(H[a,m,\alpha;\widehat\eta]-b_0\right)u_0
+\tau\left\{\left(H[a,m,\alpha;\widehat\eta]-b_0\right)u_1-(b_1+\zeta)u_0\right\}+\tau^2R,
\end{eqnarray*}
where
$$\zeta(t)=\left\{\begin{array}{l}-1;\quad{\rm if}\quad t>0,\\
a;\quad {\rm if}\quad t<0,\end{array}\right.$$ and
$$R=\left\{\begin{array}{l}
(b_1-1)u_1,\quad{\rm if}\quad t>0,\\
(b_1+a)u_1,\quad{\rm if}\quad t<0. \end{array}\right.$$ We choose
$b_0,b_1, u_0,u_1$ in the following way~:
\begin{eqnarray*}
&&b_0=\mu^{(1)}(a,m,\alpha;\widehat\eta),\quad u_0=f_{\alpha,\widehat\eta}^{a,m},\\
&&b_1=\int_{\mathbb R_+}|u_0|^2dt-a\int_{\mathbb R_-}|u_0|^2dt,\quad
u_1=\mathcal R[a,m,\alpha;\widehat\eta]g,
\end{eqnarray*}
where the function $g\in u_0^\bot$ is defined by,
$$g=\left\{
\begin{array}{l}
(b_1-1)u_0;\quad if\quad t>0,\\
(b_1+a)u_0;\quad if\quad t<0.
\end{array}
\right.
$$
Notice that $u:=u_0+\tau u_1$ is in the domain of the operator
$H[a,m,\widehat\alpha;\widehat\eta]$. Therefore, the constructions
above together with the spectral theorem yields the existence of an
eigenvalue $\widetilde \mu$ of $H[a,m,\widehat\alpha;\widehat\eta]$
that satisfies~:
$$\left|\widetilde \mu-\mu_1(a,m,\alpha;\widehat\eta)-b_1\tau\right|\leq
C\tau^2.$$ By comparing the quadratic forms
$Q[a,m,\widehat\alpha;\widehat\eta]$ and
$Q[a,m,\alpha;\widehat\eta]$, we obtain, thanks to the min-max
principle,
$$\left|\mu_2(a,m,\widehat\alpha;\widehat\eta)-\mu_2(a,m,\alpha;\widehat\eta)\right|\leq
C\tau.$$ Therefore, the only possible choice of $\widetilde\mu$ is
$\widetilde\mu=\mu_1(a,m,\widehat\alpha;\widehat\eta)$. This
achieves the proof of the lemma.
\hfill$\Box$\\

We determine in the next lemma a useful `key' estimate of
$|\widehat\eta-\eta|$.
\begin{lem}\label{II-|etan-eta|}
Let $\alpha$ be as in (\ref{II-alp-am}). There exist constants
$C,\epsilon>0$ such that, if $|\widehat\alpha-\alpha|\leq\epsilon$,
then we have~:
\begin{equation}\label{II-eq-|etan-eta|}
|\widehat\eta-\eta|\leq C|\widehat\alpha-\alpha|.
\end{equation}
Here $\eta,\widehat\eta$ are introduced in (\ref{II-eta}).
\end{lem}
\paragraph{\bf Preuve.}
We denote by~:
$$
\widehat\gamma:=\gamma(a,m,\widehat\alpha;\widehat\eta),\quad
\gamma:=\gamma(a,m,\alpha;\eta),
$$
where $\gamma(\cdot)$ is defined in (\ref{Sec2,deGe}). As
$\widehat\eta$ (respectively $\eta$) is a critical point of the
function $\mu_1(a,m,\widehat\alpha;\cdot)$ (respectively of
$\mu_1(a,m,\alpha;\cdot)$, we get, thanks to formula
(\ref{IIderVP-eq})~:
$$\widehat\gamma^2=\frac{a+1}{m-1}\widehat\alpha-\frac1m\widehat\eta^2,\quad
\gamma^2=\frac{a+1}{m-1}\alpha-\frac1m\eta^2,$$ and consequently,
\begin{equation}\label{II-(1)}
\widehat\gamma^2-\gamma^2=\frac{a+1}{m-1}(\widehat\alpha-\alpha)-\frac1m(\widehat\eta^2-\eta^2).
\end{equation}
Writing Taylor's formula up to the order $1$ of the function
$\xi\mapsto\gamma(a,m,\alpha;\xi)$ near $\eta$, we obtain,
$$
\widehat\gamma=\gamma+c_1(\widehat\eta-\eta)+\mathcal
O(|\widehat\eta-\eta|^2),
$$
where $c_1=\gamma'(\eta)$ is given by, thanks to
(\ref{II-der-gam(xi)}),
$$c_1=\gamma^2-\eta^2+\alpha.$$
Substituting in (\ref{II-(1)}), we obtain~:
$$\left(c_1\gamma+\frac\eta{m}\right)(\widehat\eta-\eta)+\mathcal
O(|\widehat\eta-\eta|^2)=\mathcal O(|\widehat\alpha-\alpha|).$$
Since $\eta$ is a non-degenerate minimum of
$\mu_1(a,m,\alpha;\cdot)$, then, thanks to (\ref{II-der-mu(xi)}),
$$c_1\gamma+\frac\eta{m}>0.$$
This achieves the proof of the lemma. \hfill$\Box$\\

Lemmas~\ref{II-mu-th'} and \ref{II-|etan-eta|} give now the
following theorem, thanks to the regularity in Theorem~\ref{reg}.

\begin{thm}\label{II-mu-th'-1}
There exist constants $C,\epsilon>0$ such that, if
$|\widehat\alpha-\alpha|\leq\epsilon$, we have,
\begin{equation}\label{II-mu-th'-2}
\left|\mu_1(a,m,\widehat\alpha;\widehat\eta)-b_1(a,m)(\widehat\alpha-\alpha)\right|\leq
C|\widehat\alpha-\alpha|^2. \end{equation}
\end{thm}

\subsection{A lower bound of $\mu_1(\mathcal
H_{h,\beta,\xi}^{a,m,\widehat\alpha})$} We start by the following
`rough' localization of the spectrum of $\mathcal
H_{h,\beta,\xi}^{a,m,n}$.
\begin{prop}\label{II-loc-sp}
Given $a,m,M>0$, there exist constants $C,h_0>0$ such that
$$\forall \beta\in]-M,M[,\quad \forall\xi\in\mathbb R,\quad \forall
h\in]0,h_0],$$ we have,
\begin{equation}\label{II-4.2.38}
\left|\mu_j(\mathcal
H_{h,\beta,\xi}^{a,m,\widehat\alpha})-\mu_j(\mathcal
H_{0,\xi}^{a,m,\widehat\alpha})\right|\leq Ch^{2\delta-1/2}
\left(1+\mu_j(\mathcal H_{0,\xi}^{a,m,\widehat\alpha})\right),
\end{equation}
where the operator $\mathcal H_{0,\xi}^{a,m,\widehat\alpha}$ is
defined by~:
$$D(\mathcal H_{0,\xi}^{a,m,\widehat\alpha}):=D(\mathcal
H_{h,\beta,\xi}^{a,m,\widehat\alpha}),\quad
H_{0,\xi}^{a,m,\widehat\alpha}=H[a,m,\widehat\alpha;\xi].$$
\end{prop}
The estimate (\ref{II-4.2.38}) is obtained by first  comparing the
corresponding quadratic forms and then by applying  the
min-max principle. Notice that the min-max principle gives also,
thanks to the inclusion of the form domains,
\begin{equation}\label{II-4.2.45}
\forall j\in\mathbb N,\quad\mu_j(\mathcal
H_{0,\xi}^{a,m,\widehat\alpha})\geq\mu_j(a,m,\widehat\alpha;\xi),
\end{equation}
where $\mu_j(a,m,\widehat\alpha;\xi)$ is the increasing sequence of
eigenvalues of the operator $H[a,m,\widehat\alpha;\xi]$.\\
Since $\widehat\eta$ is a non-degenerate minimum of
$\mu^{(1)}(a,m,\widehat\alpha;\cdot)$, we get the following lower
bound of $\mu_1(\mathcal H_{h,\beta,\xi}^{a,m,\widehat\alpha})$,
when $\xi$ is not very near $\widehat\eta$.

\begin{prop}\label{II-xi-eta>}
Given $a,m,M>0$, there exist constants $C,\zeta,h_0>0$ such that~:
\begin{equation}\label{II-xi-hyp}
\forall \beta\in]-M,M[,\quad \forall \xi~:~
|\xi-\widehat\eta|\geq\zeta h^{\delta-1/4},\quad \forall
h\in]0,h_0],
\end{equation}
we have,
\begin{equation}\label{II-4.2.46}
\mu_1(\mathcal H_{h,\beta,\xi}^{a,m,\widehat\alpha})\geq
\mu_1(a,m,\widehat\alpha;\widehat\eta)+Ch^{2\delta-1/2}.
\end{equation}
\end{prop}
\paragraph{\bf Proof.}
Recalling the localization of the spectrum in (\ref{II-4.2.38}) and
(\ref{II-4.2.45}), it is sufficient to prove the existence of
$\zeta>0$ and $h_0$ such that, under the hypothesis
(\ref{II-xi-hyp}), we have,
$$\mu_1(a,m,\widehat\alpha;\xi)\geq
\mu_1(a,m,\widehat\alpha;\widehat\eta)+h^{2\delta-1/2}.$$ Using
Taylor's formula up to the order $2$, we get constants
$\theta,C_0>0$ such that~:
\begin{eqnarray}\label{Tay-2-ndeg}
\mu_1(a,m,\widehat\alpha;\xi)&=&\mu_1(a,m,\widehat\alpha;\widehat\eta)+
\frac{|\xi-\widehat\eta|^2}2\partial_\xi^2\mu_1(a,m,\widehat\alpha;\xi)_{{\xi=\widehat\eta}}\\
&&-C_0|\xi-\widehat\eta|^3,\quad\forall
\xi\in]\widehat\eta-\theta,\widehat\eta+\theta[.\nonumber
\end{eqnarray}
The constant $C_0$ is uniform with respect to $\widehat\alpha$,
thanks to the regularity
in Theorem~\ref{reg}.\\
Since
$\partial_\xi^2\mu^{(1)}(a,m,\widehat\alpha;\xi)_{{\xi=\widehat\eta}}>0$
(cf. Theorem~\ref{IIRe-Gen-Th}), there exist constants $C_0'>0$ such
that we can rewrite (\ref{Tay-2-ndeg}) as~:
$$
\mu_1(a,m,\widehat\alpha;\xi)\geq\mu_1(a,m,\widehat\alpha;\widehat\eta)+
C_0'|\xi-\widehat\eta|^2,\quad\forall
\xi\in]\widehat\eta-\theta,\widehat\eta+\theta[.
$$
We choose $\zeta>0$ in such a way that $C_0'\zeta>1$. We then obtain
obtain for $\zeta h^{\delta-1/2}\leq|\xi-\widehat\eta|<\theta$,
$$\mu_1(a,m,\widehat\alpha;\xi)\geq
\mu_1(a,m,\widehat\alpha;\widehat\eta)+h^{2\delta-1/2}.$$ If
$|\xi-\widehat\eta|\geq\theta$, there exists, thanks to the
variation of $\mu_1(a,m,\widehat\alpha;\cdot)$, a constant
$\epsilon_\theta>0$ such that,
$$\mu_1(a,m,\widehat\alpha;\xi)\geq\mu_1(a,m,\widehat\alpha;\widehat\eta)+\epsilon_\theta.$$
It is sufficient now to choose $h_0$ in such a way that
$h_0^{2\delta-1/2}<\epsilon_\theta$.\hfill$\Box$\\

We suppose now that $|\xi-\widehat\eta|\leq\zeta h^{2\delta-1/2}$.
In this case we follow the general technique initiated
in~\cite[Section 11]{HeMo3} to construct a formal asymptotic
expansion for $\mu_1(\mathcal H_{h,\beta,\xi}^{a,m,\widehat\alpha})$
in powers of
$(\xi-\widehat\eta)$.\\
We look for a formal solution
$\left(\mu,f^{a,m,\widehat\alpha}_{h,\beta,\xi}\right)$ of the
following eigenvalue problem\footnote{For two functions $f(h),g(h)$,
we use the
  notation $f\sim g$ if $\lim_{h\to0}([f(h)]/[g(h)]=1$.},
\begin{equation}\label{II-PSpFML}
\left\{\begin{array}{l}
\left(H_{h,\beta,\xi}-\widehat\alpha\right)f_{h,\beta,\xi}^{a,m,\widehat\alpha}\sim
\mu f_{h,\beta,\xi}^{a,m,\widehat\alpha},\quad\text{in }\mathbb R_+,\\
\\
\left(\frac1mH_{h,\beta,\xi}+a\widehat\alpha\right)f_{h,\beta,\xi}^{a,m,\widehat\alpha}\sim
\mu f_{h,\beta,\xi}^{a,m,\widehat\alpha},\quad\text{in }\mathbb R_-,\\
\\
(f^{a,m,\widehat\alpha}_{h,\beta,\xi})'(0_+)=\frac1m(f^{a,m,\widehat\alpha}_{h,\beta,\xi})'(0_-)
\text{ in }\mathbb R,
\end{array}\right.
\end{equation}
in the form~:
\begin{eqnarray}\label{II-VPFML}
&&\hskip-0.5cm\mu=d_0+d_1\left(\xi-\widehat\eta\right)
+d_2\left(\xi-\widehat\eta\right)^2+d_3 h^{1/2},\\
&&\hskip-0.5cm
f_{h,\beta,\xi}^{a,m,\widehat\alpha}=u_0+\left(\xi-\widehat\eta\right)u_1
+\left(\xi-\widehat\eta\right)^2u_2+h^{1/2}u_3,\label{II-FPFML}
\end{eqnarray}
where the coefficients $d_0,d_1,d_2,d_3$ and the functions
$u_0,u_1,u_2,u_3$ are to be determined. We expand the operator
$H_{h,\beta,\xi}$ in powers of $(\xi-\xi(\widetilde\eta))$ and then
we identify the coefficients in (\ref{II-PSpFML}) of the terms of
orders $(\xi-\xi(\widetilde\eta))^j$ ($j=0,1,2$) and $h^{1/2}$. We
then obtain the following `leading order' equations,
\begin{eqnarray}
&&\hskip-0.7cm\left(H[a,m,\widehat\alpha;\widehat\eta]-d_0\right)u_0=0,\label{II-cofd0}\\
&&\hskip-0.7cm\left(H[a,m,\widehat\alpha;\widehat\eta]-d_0\right)u_1=g_1,\label{II-cofd1}\\
&&\hskip-0.7cm\left(H[a,m,\widehat\alpha;\widehat\eta]-d_0\right)u_2=g_2,\label{II-cofd2}\\
&&\hskip-0.7cm\left(H[a,m,\widehat\alpha;\widehat\eta]-d_0\right)u_3=g_3,\label{II-cofd3}
\end{eqnarray}
where the function $g_1,g_2$ and $g_3$ are defined by~:
\begin{equation}\label{II-g1}
g_1(t)=\left\{\begin{array}{l} 2(t-\widehat\eta)u_0,\quad{\rm if}\quad t>0,\\
\frac2m(t-\widehat\eta)u_0,\quad{\rm if}\quad t<0,
\end{array}\right.
\end{equation}
\begin{equation}\label{II-g2}
g_2(t)=\left\{\begin{array}{l}
2\left[(t-\widehat\eta)u_1+d_1\right]+(d_2-1)u_0,
\quad{\rm if}\quad t>0,\\
2\left[\frac1m(t-\widehat\eta)+d_1\right]u_1+(d_2-\frac1m)u_0,\quad
{\rm if}\quad t<0,
\end{array}\right.
\end{equation}
and
\begin{equation}\label{II-g3}
g_3(t)=\left\{\begin{array}{l}
-\left[\beta\left(\partial_t+\left(t-\widehat\eta\right)^3-
\widehat\eta^2\left(t-\widehat\eta\right)\right)
-d_3\right]u_0,\quad
{\rm if}\quad t>0,\\
-\left[\frac1m\beta\left(\partial_t+\left(t-\widehat\eta\right)^3
-\widehat\eta^2\left(t-\widehat\eta\right)\right)
-d_3\right]u_0,\quad {\rm if}\quad t<0.
\end{array}\right.
\end{equation}
We recall that (cf. (\ref{II-M1}))~:
\begin{equation}\label{II-M1'}
\int_{\mathbb R_+}(t-\widehat\eta)|\widehat
f(t)|^2dt+\frac1m\int_{\mathbb R_-}(t-\widehat\eta)|\widehat
f(t)|^2dt=0.
\end{equation}
Then, thanks to (\ref{II-M1'}), we can choose $d_0,d_1,d_2, d_3$ and
the functions $u_0,u_1,u_2,u_3$ in the following way~:
\begin{eqnarray}\label{II-coef}
&&d_0=\mu_1(a,m,\widehat\alpha;\widehat\eta),\quad u_0=\widehat f,\label{II-d0}\\
&& d_1=0,\quad
u_1=2\mathcal R[a,m,\widehat\alpha;\widehat\eta]g_1,\label{II-d1}\\
&&d_2=\int_{\mathbb
R_+}\left(u_0-2(t-\widehat\eta)u_1\right)u_0\,dt+\frac1m\int_{\mathbb
R_-}\left(u_0-2(t-\widehat\eta)\right)u_0\,dt,\label{II-d2}\\
&&u_2=2\mathcal R[a,m,\widehat\alpha;\widehat\eta]g_2,\label{II-u2}\\
&& d_3=\beta\left(\int_{\mathbb
R_+}u_0\cdot\left(\partial_t+(t-\widehat\eta)^3\right)u_0\,dt
+\frac1m\int_{\mathbb
R_-}u_0\cdot\left(\partial_t+(t-\widehat\eta)^3\right)u_0\,dt\right),\label{II-d3}\\
&& u_3=\beta \mathcal
R[a,m,\widehat\alpha;\widehat\eta]g_3.\label{II-u3}
\end{eqnarray}
An integration by parts yields~:
\begin{equation}\label{II-d3=C1}
d_3=\beta\, \widehat {\mathcal C}_1(a,m),
\end{equation}
where
\begin{equation}\label{II-C1}
\widehat {\mathcal C}_1(a,m)=\int_{\mathbb
R_+}(t-\widehat\eta)^3|\widehat f(t)|^2dt+\frac1m\int_{\mathbb
R_-}(t-\widehat\eta)^3|\widehat f(t)|^2dt
+\frac12\left(1-\frac1m\right)|\widehat f(0)|^2.
\end{equation}
We define now the following quasi-mode~:
\begin{equation}\label{II-f-nh}
\widehat
f_h(t)=\chi\left(h^{-\delta+1/2}t\right)f_{h,\beta,\xi}^{a,m,\widehat\alpha}(t),
\end{equation}
where $\chi$ is a cut-off function supported in $]-1,1[$. Since the
function $\widehat f$ decays exponentially at infinity, we get~:
$$\left|\|\widehat f_h\|^2_{L^2(\mathbb R^2)}-1\right|\leq
C[|\xi-\widehat \eta|+h^{1/2}],\quad \forall h\in]0,h_0].$$ We have
also, thanks to the formal calculus presented
above~((\ref{II-PSpFML})-(\ref{II-u3})),
\begin{eqnarray*}
&&\left|\left(\mathcal
H_{h,\beta,\xi}^{a,m,\widehat\alpha}-[d_0+d_2(\xi-\widehat\eta)^2+d_3h^{1/2}]\right)\widehat f_h\right|\\
&&\leq C[h^{1/2}|\xi-\widehat\eta|+h^{1/2+\delta}],\quad \forall
h\in]0,h_0].
\end{eqnarray*}
By the spectral Theorem, we get an eigenvalue $\lambda(\mathcal
H_{h,\beta,\xi}^{a,m,\widehat\alpha})$ of $\mathcal
H_{h,\beta,\xi}^{a,m,\widehat\alpha}$ such that~:
\begin{eqnarray*}
&&\left|\lambda(\mathcal
H_{h,\beta,\xi}^{a,m,\widehat\alpha})-[d_0+d_2(\xi-\widehat\eta)^2
+d_3h^{1/2}]\right|\\
&&\leq C[h^{1/2}|\xi-\widehat\eta|+h^{1/2+\delta}],\quad \forall
h\in]0,h_0].
\end{eqnarray*}
The localization of the spectrum of $\mathcal
H_{h,\beta,\xi}^{a,m,\widehat\alpha}$ in Proposition~\ref{II-loc-sp}
together with the lower bound (\ref{II-4.2.45}) shows that the only
possible choice of $\lambda(\mathcal
H_{h,\beta,\xi}^{a,m,\widehat\alpha})$ is
$$\lambda(\mathcal H_{h,\beta,\xi}^{a,m,\widehat\alpha})=\mu_1(\mathcal
H_{h,\beta,\xi}^{a,m,\widehat\alpha}).$$ Therefore, we have proved
the following lemma.

\begin{lem}\label{II-lem4.2.15}
Given $a,m,M>0$, there exist constants $C,h_0>0$ such that,
$$\forall \beta\in]-M,M[,\quad \forall\xi\text{ \rm s.t.}\quad
|\xi-\widehat\eta|\leq\zeta h^{\delta-1/4},\quad\forall
h\in]0,h_0],$$ we have~:
\begin{eqnarray*}
&&\left|\mu_1(\mathcal
H_{h,\beta,\xi}^{a,m,\widehat\alpha})-[d_0+d_2(\xi-\widehat\eta)^2+d_3h^{1/2}]\right|
\leq Ch^{1/4+\delta},\quad \forall h\in]0,h_0].
\end{eqnarray*}
\end{lem}

The next lemma permits to deduce that $d_2>0$.

\begin{lem}\label{II-lem4.2.16}
Under the hypothesis of Lemma~\ref{II-lem4.2.15}, we have~:
$$
d_2=\frac12\partial_\xi^2\mu_1(a,m,\widehat\alpha;\xi)_{|_{\xi=\widehat\eta}}.$$
\end{lem}
\paragraph{\bf Proof.}
Notice that, by construction of
$f_{h,\beta,\xi}^{a,m,\widehat\alpha}$, one has,
$$\left\|\left(H[a,m,\widehat\alpha;\xi]-[d_0+d_2(\xi-\widehat\eta)^2]\right)
f_{h,\beta,\xi}^{a,m,\widehat\alpha}\right\|_{L^2(\mathbb R^2)}\leq
C|\xi-\widehat\eta|^2.$$ By the spectral theorem, we get,
$$\left|\mu_1(a,m,\widehat\alpha;\xi)-[d_0+d_2(\xi-\widehat\eta)^2]\right|\leq
C|\xi-\widehat\eta|^2.$$ Comparing the above expansion with that
obtained after writing Taylor's Formula for
$\mu_1(a,m,\widehat\alpha;\xi)$ up to the order $2$, one gets the
result
of the lemma.\hfill$\Box$\\

Using  Proposition~\ref{II-xi-eta>} and Lemmas~\ref{II-lem4.2.15}
and \ref{II-lem4.2.16}, we get finally the following theorem.

\begin{thm}\label{II-tp-mop-M}
Under the hypothesis of Theorem~\ref{II-thmF}, given $a,m,M>0$,
there exist constants $C,h_0>0$ such that~:
$$\forall\beta\in]-M,M[,\quad\forall \xi\in\mathbb R,\quad\forall
h\in]0,h_0],$$ we have~:
\begin{equation}\label{II-tp-eq}
\mu_1(\mathcal
H_{h,\beta,\xi}^{a,m,\widehat\alpha})\geq\mu_1(a,m,\widehat\alpha;\widehat\eta)+
\widehat{\mathcal C}_1(a,m)h^{1/2}-Ch^{1/4+\delta}.
\end{equation}
\end{thm}

Using Theorem~\ref{II-tp-mop-M}, the regularity of the eigenfunction
in Theorem~\ref{reg}, and
Lemmas~\ref{II-mu-th'} and \ref{II-|etan-eta|}, we achieve the proof
of Theorem~\ref{II-thmF}.

\section{Estimates for the bottom of the
spectrum}\label{k2-Est-btSp}

Let us denote by $\mathcal P[\kappa,H]$ the self-adjoint operator
associated to the quadratic form (\ref{qf}) by Friedrichs' theorem.
In this section, we estimate the bottom of the spectrum of $\mathcal
P[\kappa,H]$ in the regime $\kappa,H\to+\infty$ and we prove
Theorem~\ref{k2Th1}.\\
We introduce the following parameter,
\begin{equation}\label{epsilon}
\varepsilon=\frac1{\sqrt{\kappa H}},
\end{equation}
which will be small in our analysis.

We start with a `rough' estimate of $\mu^{(1)}(\kappa,H)$ which
gives an alternative characterization of the upper critical field
$H_{C_3}(a,\widetilde m;\kappa)$.

\begin{prop}\label{Min-max-U}
Given $a,m>0$, there exist constants $C,\varepsilon_0>0$ such that,
when $\varepsilon\in]0,\varepsilon_0]$, we have,
\begin{eqnarray}\label{Min-max-U-eq}
\hskip0.5cm
-C\varepsilon+\min\left(\Theta_0-\frac{\kappa}{H},\frac1m\Theta_0+a\frac{\kappa}H\right)
&\leq& \varepsilon^2\mu^{(1)}(\kappa,H)\\
&\leq&
\min\left(1-\frac{\kappa}{H},\frac1m+a\frac{\kappa}H\right)+C\varepsilon.
\nonumber
\end{eqnarray}
Moreover, there exists a constant $\kappa_0>0$ such that, if
$\kappa\geq\kappa_0$, then,
\begin{equation}\label{HC3=H*}
H_{C_3}(a,m;\kappa)=\min\{H>0;\quad \mu^{(1)}(\kappa,H)=0\}.
\end{equation}
\end{prop}
\paragraph{\bf Proof.}\ \\
Using the min-max principle, we get,
\begin{eqnarray}\label{II-H*=O(kappa)}
&&\min\left(\varepsilon^{2}\mu^N(\varepsilon;\Omega)-\frac{\kappa}H,
\frac{\varepsilon^{2}}{m}\mu^N(\varepsilon;\Omega^c)+a\frac\kappa{H}\right)\nonumber\\
&&\leq
\varepsilon^2\mu^{(1)}(\kappa,H)\\
&&\leq
\min\left(\varepsilon^{2}\mu^D(\varepsilon;\Omega)-\frac{\kappa}H,
\frac{\varepsilon^{2}}{m}\mu^D(\varepsilon;\Omega^c)+a\frac\kappa{H}\right),\nonumber
\end{eqnarray}
where $\mu^N(\varepsilon;\cdot)$, $\mu^D(\varepsilon;\cdot)$ are
defined in (\ref{mu-N,D}).  Estimate (\ref{Min-max-U-eq}) follows
now from
Proposition~\ref{HeMo3-est-ND}.\\
We prove now (\ref{HC3=H*}). We define,
\begin{equation}\label{H*}
H_*(\kappa)=\min\{H>0;\quad \mu^{(1)}(\kappa,H)=0\}.
\end{equation}
By definition, $H_{C_3}(a,m;\kappa)\leq H_*(\kappa)$. By
(\ref{II-H*=O(kappa)}), $H_*(\kappa)$ has the order of $\kappa$ (as
$\kappa\to+\infty$). Suppose by contradiction that there exist
$H\in]0,H_*(\kappa)[$ such that
$$\mu^{(1)}(\kappa,H)>0.$$
Let $H_1=\sqrt{\kappa}$ and choose $\kappa_0$ large enough, thanks
to the upper bound in (\ref{Min-max-U-eq}), such that,
$$\mu^{(1)}(\kappa,H_1)<0,\quad \forall \kappa\geq\kappa_0.$$
As the function $H\mapsto\mu^{(1)}(\kappa,H)$ is continuous, we get
by the Intermediate Value Theorem a contradiction
to the definition of $H_*(\kappa)$.\hfill$\Box$\\

\begin{rem}\label{H*-equ}
Let
\begin{equation}\label{alp(kap)}
\alpha(\kappa):=\frac{\kappa}{H_*(\kappa)}.
\end{equation}
Then, thanks to (\ref{Min-max-U-eq}), there exist positive constants
$C,C',\kappa_0$ such that,
\begin{equation}\label{alp(kap)-est}
C'\leq \alpha(\kappa)\leq C,\quad\forall \kappa\geq\kappa_0.
\end{equation}
Moreover, thanks to (\ref{HC3=H*}), it is sufficient to calculate
$\displaystyle\lim_{\kappa\to+\infty}\alpha(\kappa)$ in order to
prove Theorem~\ref{k2Th1}.
\end{rem}

We modify our notation slightly by redefining the parameter
$\varepsilon$,
\begin{equation}\label{epsilon-new}
\varepsilon=\frac1{\sqrt{\kappa H_*(\kappa)}}.
\end{equation}

\begin{prop}\label{UB-mu1}(Lower bound)\ \\
There exist constants $C,\varepsilon_0>0$ such that,
\begin{equation}\label{UB-eq1}
\inf_{\xi\in\mathbb
R}\mu_1\left(a,m,\alpha(\kappa);\xi\right)+C\varepsilon\geq0,\quad
\forall\varepsilon\in]0,\varepsilon_0],
\end{equation}
where the function $\mu_1(\cdot)$ is defined in (\ref{IIpVP}).
\end{prop}
\paragraph{\bf Proof.}
If $m\leq1$, then, thanks to Theorem~\ref{min-thm1},
$$\inf_{\xi\in\mathbb R}\mu_1\left(a,m,\alpha(\kappa);\xi\right)
=1-\alpha(\kappa).$$ In this case,
(\ref{UB-eq1}) comes from the upper bound in (\ref{Min-max-U-eq}).\\
Let us suppose now that $m>1$. If  the function
$$\xi\mapsto\mu_1\left(a,\frac{m}\mu,\alpha(\kappa);\xi\right)$$
does not attain its minimum, we have then nothing to prove, thanks
to (\ref{lim-pm-inf}) and (\ref{Min-max-U-eq}).\\
Suppose now that the function
$\xi\mapsto\mu_1\left(a,m,\alpha(\kappa);\xi\right)$ attains its
minimum at a point $\eta=\eta\left(a,m,\alpha(\kappa)\right)$.
Notice that $\eta(\kappa)$ is bounded, thanks to (\ref{|xi|<M}) and
(\ref{alp(kap)-est}). Let $x_0$ be an arbitrary point of
$\partial\Omega$. Using the boundary coordinates $(s,t)$ introduced
in Subsection~\ref{sbS-BC}, we construct a trial function $u(s,t)$
supported near $x_0$. We can suppose that $x_0=(0,0)$ in the $(s,t)$
coordinate system. Let $\chi$ be a cut-off function such that,
\begin{equation}\label{cu-off-Sec5}
{\rm supp}\,\chi\subset]-t_0,t_0[,\quad0\leq\chi\leq1,\quad
\chi=1\text{ in }[-t_0/2,t_0/2],
\end{equation}
where the constant $t_0$ is the geometric constant introduced in
Subsection~\ref{sbS-BC}. Let us consider another function $f$ such
that,
\begin{equation}\label{f-Sec5}
f\in C_0^\infty(]-1/2,1/2[),\quad \|f\|_{L^2(\mathbb R)}=1.
\end{equation}
We denote also by $f_{\alpha,\eta}$ the first (positive)
eigenfunction of the operator $H[a,\frac{m}\mu,\alpha(\kappa);\eta]$
whose $L^2$-norm in $\mathbb
R$ is equal to $1$.\\
We define $u(s,t)$ by,
\begin{equation}\label{IIqm+}
u(s,t)= a^{-1/2}(s,t)\varepsilon^{-3/4}\exp\left(-\frac{i\eta
s}{\varepsilon}\right)f_{\alpha,\eta}\left(\varepsilon^{-1}t\right)\chi(t)\times
f(\varepsilon^{-1/2}s).
\end{equation}
We recall the definition of $a(s,t)$ in (\ref{Jac}). Since the
function $f_{\alpha,\eta}$ decays exponentially at infinity, we get,
thanks to (\ref{nostco}),
\begin{equation}\label{IIN=1}
1-C\exp\left(-\frac1\varepsilon\right)\leq\|u\|^2_{L^2(\mathbb
R^2)}\leq1.
\end{equation}
Working with the gauge introduced in Proposition~\ref{Agd1}, we get,
thanks to the change of variable formulas (\ref{qfstco}) and
(\ref{qfstco'})~:
\begin{equation}\label{IIfq-est}
\left|\varepsilon^2\mathcal
Q[\kappa,H_*](u)-Q\left[a,\frac{m}\mu,\alpha(\kappa);\eta\right](f_{\alpha,\eta})\right|\leq
C\varepsilon.
\end{equation}
By our choice of  $f_{\alpha,\eta}$ and since $\mu_1(\kappa,H_*)=0$,
the application of the min-max principle achieves the proof.
\hfill$\Box$\\

\begin{prop}\label{LB-mu1}(Upper bound)\ \\
There exist constants $C,\varepsilon_0>0$ such that, if
$\varepsilon\in]0,\varepsilon_0]$, then,
\begin{equation}\label{LB-kH-inftyEq}
\left(1-C\varepsilon^{1/2}\right)\inf_{\xi\in\mathbb R}\mu_1
\left(a,m,\alpha(\kappa);\xi\right) -C\varepsilon^{1/2}\leq0,
\end{equation}
where $\varepsilon$ has been introduced in (\ref{epsilon-new}).
\end{prop}
\paragraph{\bf Proof.}
We follow the technique of Helffer-Morame~\cite[Subsection
6.3]{HeMo3} and we localize by means of a partition of unity to
compare with the model operator.\\
Let $0<\rho<2$. Consider the partition of unity
$(\chi_j^\varepsilon)$ defined by Proposition~\ref{P-O-U}. We have
the following decomposition formula, thanks to (\ref{IMS-f}),
\begin{equation}\label{II-IMS}
Q[\kappa,H_*](u)=\sum _jQ[\kappa,H_*](\chi _j^\varepsilon u) -\sum
_j\|\,|\nabla\chi _j^\varepsilon|\,u\|^2,\quad \forall u\in \mathcal
H^1_{\varepsilon^{-2}\mathbf F}(\mathbb R^2),
\end{equation}
where $\|.\|$ denotes the $L^2$-norm in  $\mathbb R^2$.\\
The alternative appearing in (\ref{P-O-U-eq3}) permits one to
decompose the above sum
 in the following form~:
$$\sum=\sum_{{\rm int}}+\sum_{{\rm ext}}+\sum_{{\rm bnd}},$$
where the summation over `int' means that we sum over the $j$'s such
that $\chi_j^{\varepsilon}$ is supported in $\Omega$, that over
`ext' means that $\chi_j^\varepsilon$ is supported in
$\overline{\Omega}^c$ and that over `bnd' means that the support of
$\chi_j^\varepsilon$ meets the
boundary $\partial\Omega$.\\
We have to bound from below each of the terms on the right hand side
of (\ref{II-IMS}). By (\ref{P-O-U-eq2}), we can estimate the
contribution of the last term in (\ref{II-IMS})~:
\begin{equation}\label{IIclb2}
\sum _j\|\,|\nabla\chi _j^\varepsilon|\,u\|^2\leq C\zeta
_0^{-2}\varepsilon^{-2\rho}\|u\|^2.
\end{equation}
Using (\ref{HeMo3-int}) and our choice of $\mathbf F$ in (\ref{F}),
we get,
\begin{eqnarray}
&&\sum_{\rm int}\mathcal Q[\kappa,H_*](\chi_j^\varepsilon u)\geq
\varepsilon^{-2}
\sum_{\rm int}\left(1-\frac{\kappa}{H_*}\right)\|\chi_j^\varepsilon u\|^2,\label{II-int1}\\
&&\sum_{\rm ext}\mathcal Q[\kappa,H_*](\chi_j^\varepsilon u)\geq
\varepsilon^{-2} \sum_{\rm
ext}\left(\frac1{m}+a\frac{\kappa}{H_*}\right)\|\chi_j^\varepsilon
u\|^2.\label{II-ext1}
\end{eqnarray}
We have only now to bound from below $\sum_{\rm bnd}\mathcal
Q[\kappa,H_*](\chi_j^\varepsilon u)$. In the support of
$\chi_j^\varepsilon$, we choose the gauge from
Proposition~\ref{Agd1}. Proposition~\ref{HeMo3-Bprop} yields now the
existence of constants $C,\varepsilon_0>0$ depending only on
$\Omega$, and a function $\phi_j\in H^1_{\rm loc}(\mathbb R^2)$ such
that, for any $\theta>0$ and $\varepsilon\in]0,\varepsilon_0]$,
\begin{eqnarray}\label{II-bnd-est}
&&\\
&&\sum_{\rm bnd}\mathcal Q[\kappa,H_*](\chi_j^\varepsilon u)\geq
\left(1-C\zeta_0\varepsilon^\rho-C\varepsilon^{2\theta}\right)
\sum_{\rm bnd}\mathcal Q_{\mathbb R\times\mathbb R_+}[\kappa,H_*]
\left(\exp\left(-i\frac{\phi_j}\varepsilon\right)\chi_j^\varepsilon u\right)\nonumber\\
&&\hskip3cm-C\varepsilon^{-2}
\left(\varepsilon^{4\rho-2\theta-2}+\zeta_0\varepsilon^\rho+\varepsilon^{2\theta}\right)
\sum_{\rm brd}\|\chi_j^\varepsilon u\|^2,\nonumber
\end{eqnarray}
where the quadratic form $\mathcal Q_{\mathbb R\times\mathbb
R_+}[\kappa,H_*]$ is defined by (\ref{k2HP-qf}). Notice that we have
also used the estimate (\ref{alp(kap)-est}). Formula
(\ref{II-bnd-est}) now reads, thanks to Remark~\ref{Rel-HP},
\begin{eqnarray}\label{IIest-brd}
&&\\
&&\hskip-0.5cm\sum_{\rm bnd}Q[\kappa,H_*](\chi_j^\varepsilon u)\geq
\varepsilon^{-2}\left(1-C\zeta_0\varepsilon^\rho-C\varepsilon^{2\theta}\right)\left(
\inf_{\xi\in\mathbb
R}\mu_1\left(a,m,\alpha(\kappa);\xi\right)\right)\sum_{\rm
bnd}\|\chi_j^\varepsilon
u\|^2\nonumber\\
&&\hskip2.5cm-C\varepsilon^{-2}
\left(\varepsilon^{4\rho-2\theta-2}+\zeta_0\varepsilon^\rho+\varepsilon^{2\theta}\right)
\sum_{\rm bnd}\|\chi_j^\varepsilon u\|^2.\nonumber
\end{eqnarray}
Summing up the estimates (\ref{IIclb2}), (\ref{II-int1}),
(\ref{II-ext1}) and (\ref{IIest-brd}), the decomposition formulas
(\ref{II-IMS}) read as,
\begin{eqnarray}\label{II-conc}
&&\\
\mathcal Q[\kappa,H_*](\chi_j^\varepsilon u)&\geq& \varepsilon^{-2}
\left(\sum_{\rm
int}\left(1-\frac{\kappa}{H_*}\right)\|\chi_j^\varepsilon u\|^2
+\sum_{\rm ext}\left(\frac1{m}+a\frac{\kappa}{H_*}\right)\|\chi_j^\varepsilon u\|^2\right)\nonumber\\
&&+\varepsilon^{-2}\left(1-C\zeta_0\varepsilon^\rho-C\varepsilon^{2\theta}\right)
\left(\inf_{\xi\in\mathbb R}\mu_1\left(a,m
,\alpha(\kappa);\xi\right)\right)\sum_{\rm bnd}\|\chi_j^\varepsilon u\|^2\nonumber\\
&&-C\varepsilon^{-2}\left(\varepsilon^{4\rho-2\theta-2}
+\zeta_0\varepsilon^\rho+\varepsilon^{2\theta}+\varepsilon^{2-2\rho}\right)\|u\|^2.\nonumber
\end{eqnarray}
The optimal choice of $\rho$ and $\theta$ seems to be when
$4\rho-2\theta-2=2-2\rho=2\theta$, i.e. $\rho=3/4$ and $\theta=1/4$.
With this choice, and taking $\zeta_0=1$, we get the lower bound
(\ref{LB-kH-inftyEq}) after the application of the min-max principle
and by remembering that $\mu_1(\kappa,H_*)=0$. Notice that we have
used also the following fact which results from (\ref{lim-pm-inf}),
$$\inf_{\xi\in\mathbb R}
\mu_1\left(a,m,\alpha(\kappa);\xi\right) \leq
\min\left(1-\alpha(\kappa),\frac1{m}+\alpha(\kappa)\right).$$
\hfill$\Box$\\

\paragraph{\it Proof of Theorem~\ref{k2Th1}.}
Let $\kappa_n$ be a sequence such that~:
$$\lim_{n\to+\infty}\kappa_n=+\infty,\quad \exists\,\alpha_0>0~:~\lim_{n\to+\infty}\alpha(\kappa_n)=\alpha_0.$$
Then, thanks to Propositions~\ref{UB-mu1},~\ref{LB-mu1} and the
regularity in Theorem~\ref{reg}, we get,
$$\inf_{\xi\in\mathbb
R}\mu_1\left(a,m,\alpha_0;\xi\right)=0.$$ Therefore, by
Theorem~\ref{IIestalp-rem}, we should have,
$$\alpha_0=\alpha\left(a,m\right).$$
This achieves the proof of Theorem~\ref{k2Th1}, thanks to
Remark~\ref{H*-equ}.\hfill$\Box$

\section{Existence and decay of eigenfunctions}\label{Sec-Pe}
Let us denote by $\mathcal P$ the self-adjoint operator associated
to the quadratic form (\ref{qf}) when $\kappa=H=1$. The bottom of
the essential spectrum for a Schr\"odinger operator with electric
potential is characterized by Persson's Lemma~\cite{Pe}. The proof
of Persson's Lemma in~\cite{Ag} can be imitated  so that we obtain
the following characterization of the essential spectrum of
$\mathcal P$ (see \cite[Theorem~9.4.1]{kachTh}).
\begin{lem}\label{Pe-lem}
Suppose that $\Omega\subset\mathbb R^2$ is bounded. The bottom of
the essential spectrum of $\mathcal P$ is given by,
$$\inf\,\sigma_{\rm ess}(\mathcal P)=\Sigma(\mathcal P),$$
where,
$$\Sigma(\mathcal P)=\sup_{\mathcal K\subset\Omega^c}\inf\left\{
\frac1{m}\|\nabla_{\mathbf F}\phi\|^2_{L^2(\mathbb R^2)}+a;\quad
\phi\in C_0^\infty(\Omega^c\setminus\mathcal K),\quad
\|\phi\|_{L^2(\mathbb R^2)}=1\right\},$$ and the upper bound above
is taken over all compact sets $\mathcal K$ in $\Omega^c$.
\end{lem}

Using the above characterization of the essential spectrum, one can
obtain the existence of eigenfunctions of the operator $\mathcal
P[\kappa,H]$.

\begin{prop}\label{ext-EF}
Let $H=H_{C_3}(a,m;\kappa)$. There exists a constant $\kappa_0>0$
such that, if $\kappa\geq\kappa_0$, then $\mu^{(1)}(\kappa,H)$ is an
eigenvalue of the operator $\mathcal
P[\kappa,H]$.\\
Moreover, if $m>1$, denoting by $\phi_\kappa$ a ground state of
$\mathcal P[\kappa,H]$, then $\phi_\kappa$ is exponentially
localized near the boundary in the following sense,
\begin{eqnarray}\label{exp-loc}
&&\exists\,\varepsilon_0,\delta\in]0,1],\quad\exists\, C>0,\quad
\quad \forall\,\varepsilon\in]0,\varepsilon_0],\nonumber\\
&&\left\|\exp\left(\delta\frac{{\rm
dist}(x,\partial\Omega)}{\varepsilon}\right)\phi_\kappa\right\|_{L^2(\mathbb
R^2)}\leq C\|\phi_\kappa\|^2_{L^2(\mathbb R^2)},\\
&&\left\|\exp\left(\delta\frac{{\rm
dist}(x,\partial\Omega)}{\varepsilon}\right)\phi_\kappa\right\|_{H^1(\mathbb
R^2)}\leq C\varepsilon^{-1}\|\phi_\kappa\|^2_{L^2(\mathbb
R^2)}.\nonumber
\end{eqnarray}
\end{prop}
\paragraph{\bf Proof.}
Using Lemma~\ref{Pe-lem}, we get, thanks to (\ref{HeMo3-int}),
$$\inf\sigma_{\rm ess}\left(\mathcal P[\kappa,H]\right)
\geq\kappa H\left(a\frac{\kappa}{H}+\frac{1}{m}\right).$$
Proposition~\ref{UB-mu1} gives on the other hand,
$$\mu^{(1)}(\kappa,H)\leq (\kappa H)\epsilon(\kappa),$$
where
\begin{equation}\label{epsilon(kap)}
\epsilon(\kappa):=\inf_{\xi\in\mathbb
R}\mu_1\left(a,m,\alpha(\kappa);\xi\right),
\end{equation}
and $\displaystyle\lim_{\kappa\to+\infty}\epsilon(\kappa)=0$.
Therefore, there exist $\kappa_0>0$ such that, if
$\kappa\geq\kappa_0$, then,
$$\mu^{(1)}(\kappa,H)<\inf\sigma_{\rm ess}(\mathcal P[\kappa,H]),$$
and therefore $\mu^{(1)}(\kappa,H)$ is an eigenvalue.\\
We obtain the localization of the ground states via Agmon's
technique (cf.~\cite{Ag, HeMo3}). Let us explain briefly  the
argument. Let $\Phi$ be a Lipschitz function with compact support.
An integration by parts yields the following identity,
\begin{eqnarray}\label{Agm-eq1}
&&\mathcal
Q[\kappa,H]\left(\exp\left(\frac{\Phi}{\varepsilon}\right)\phi_\kappa\right)
=\mu^{(1)}(\kappa,H)\left\|\exp\left(\frac{\Phi}{\varepsilon}\right)\phi_\kappa\right\|
^2_{L^2(\mathbb R^2)}\\
&&\hskip1cm+\varepsilon^{-2}\left\|\,|\nabla\Phi|\exp\left(\frac{\Phi}{\varepsilon}\right)\phi_\kappa\right\|
^2_{L^2(\Omega)} +\frac{\varepsilon^{-2}}{
m}\left\|\,|\nabla\Phi|\exp\left(\frac{\Phi}{\varepsilon}\right)\phi_\kappa\right\|
^2_{L^2(\Omega^c)}.\nonumber
\end{eqnarray}
We denote by,
$$u=\exp\left(\frac{\Phi}{\varepsilon}\right)\phi_\kappa,\quad
\beta=\min\left(1-\frac{\kappa}{H},\frac{1}m+a\frac\kappa{H}\right),\quad
\gamma=\max\left(1,\frac1m\right).$$ Using the lower bound for
$\mathcal Q[\kappa,H]$ in (\ref{II-conc}) (with $\rho=1$ and
$\theta=1/2$) and the upper bound for $\mu^{(1)}(\kappa,H)$ in
(\ref{UB-eq1}), we rewrite (\ref{Agm-eq1}) as follows,
\begin{eqnarray}\label{Agm-eq2}
&&\left(\beta-\epsilon(\kappa)-C\zeta_0-\gamma\|\nabla\Phi\|^2_{L^\infty(\mathbb
R^2)} -C\varepsilon^{1/2}\right)\sum_{{\rm int},\rm
ext}\|\chi_j^\varepsilon u\|^2\\
&&\leq \left(\gamma\|\nabla\Phi\|^2_{L^\infty(\mathbb
R^2)}+C\zeta_0^{-2}+C\varepsilon^{1/2}\right)\sum_{\rm
bnd}\|\chi_j^\varepsilon u\|^2,\nonumber
\end{eqnarray}
where, thanks to the choice of $\rho$, each $\chi_j^\varepsilon$ is
supported in a disk of radius $\zeta_0\varepsilon$.\\
Given an integer $N$, we choose
$$\Phi=\delta\chi\left(\frac{|x|}N\right)\Phi_0(x),$$ where
$\delta\in]0,1]$ is to be determined, $\chi$ is a cut-off function,
$$0\leq\chi\leq1,\quad\chi=1\text{ in
}\left[0,\frac12\right],\quad {\rm supp}\chi\subset[0,1],$$ and
$\Phi_0$ is defined by
$$\Phi_0(x)=\max\left({\rm dist}(x,\partial\Omega),
\varepsilon\right).$$ Since $m>1$, we can choose $\zeta_0$ small
enough and $\kappa_0$ large enough such that,
$$\beta-\epsilon(\kappa)-C\zeta_0>\frac\beta2,\quad\forall\kappa\geq\kappa_0.$$
There exist also $N_0,\delta_0>0$ such that, for $N\geq N_0$ and
$\delta\in]0,\delta_0]$,
$$\frac\beta2-\gamma\|\nabla\Phi\|^2_{L^\infty(\mathbb R^2)}\geq\frac\beta4.$$
Therefor, we can rewrite (\ref{Agm-eq2}) as,
$$\int_{\mathbb R^2,|x|\leq
N}\left|\exp\left(\frac{\delta{\rm
dist}(x,\partial\Omega)}{\varepsilon}\right)\phi_\kappa\right|dx\leq
\widetilde C\|\phi_\kappa\|^2,$$ for a constant $\widetilde C>0$.
Noticing that the above estimate is  uniform with respect to $N\geq
N_0$, we get (\ref{exp-loc}) by passing to the limit $N\to+\infty$.
The bound on the $H^1$-norm follows now from
(\ref{Agm-eq1}).\hfill$\Box$\\

As a result of the decay in (\ref{exp-loc}), we get another
localization version of the ground states.
\begin{lem}\label{loc-t}
If $m>1$, then, given an integer $k\in\mathbb N$, there exist
constants $\varepsilon_k,C_k>0$ such that, for any ground state
$\phi_\kappa$ and $\varepsilon\in]0,\varepsilon_k]$, we have,
\begin{eqnarray}
&& \int_{\mathbb R^2}|t(z)|^k|\phi_\kappa(z)|^2dz\leq
C_k\varepsilon^k,\\
&&\int_{\mathbb R^2}|t(z)|^k|(\nabla-i\varepsilon^{-2}\mathbf
F)\phi_\kappa(z)|^2dz\leq C_k\varepsilon^{k-2}.
\end{eqnarray}
\end{lem}

\begin{rem}\label{conj}
It would be interesting to analyze the localization of the ground
states when $m\leq1$. It seems in this case that the ground states
should be localized in a compact subset of $\Omega$, as far as
possible of $\partial\Omega$.
\end{rem}

\section{Curvature effects and proof of
Theorem~\ref{k2Th3}}\label{k2-2Tasy} This section is devoted to the
proof of Theorem~\ref{k2Th3}. The computations that we shall carry
are similar to those carried out in
\cite[Sections~10-11]{HeMo3}.\\
Given $a>0$, we shall suppose that,
\begin{equation}\label{Hyp-m/mu}
m\geq m_0,
\end{equation}
where $m_0>1$ is defined in Theorem~\ref{IIRe-Gen-Th}. We denote by,
\begin{equation}\label{notn}
\alpha=\alpha\left(a,m\right),\quad
\widehat\alpha=\frac{\kappa}{H},\quad H=H_*(\kappa),
\end{equation}
where $H_*(\kappa)$ is defined in (\ref{H*}) and is equal to the
upper critical field for large values of $\kappa$. We shall keep the
same notations introduced in Section~\ref{Sec-MF}. In all the
proofs, $C,\varepsilon_0$ will denote generic constants independent
of $\varepsilon$ and that may change from line to line.

\begin{prop}\label{UB-cr}
With  the above hypotheses and notations, there exist constants
$C,\varepsilon>0$, such that, if $\varepsilon\in]0,\varepsilon_0]$,
then,
\begin{equation}\label{UB-cr-eq}
\varepsilon^2\mu^{(1)}(\kappa,H)\leq
\mu_1\left(a,m,\widehat\alpha;\widehat\eta\right)+\widehat{\mathcal
C}_1\left(a,m\right)(\kappa_{\rm r})_{\rm
max}\,\varepsilon+C\varepsilon^{3/2},
\end{equation}
where the function $\widehat{\mathcal C}_1(\cdot,\cdot)$ is defined
in (\ref{II-C1}).
\end{prop}
\paragraph{\bf Proof.} We construct a trial function inspired by
\cite{BeSt, HeMo3}. Let $z_0\in\partial\Omega$ be such that
$\kappa_{\rm r}(z_0)=(\kappa_{\rm r})_{\rm max}$, and suppose that
$z_0=0$ in the $(s,t)$ coordinate system introduced in
Subsection~\ref{sbS-BC}. Define the following quasi-mode,
\begin{equation}\label{II-qm-BS}
u_\varepsilon(s,t)=
\varepsilon^{-5/8}a_0^{-1/2}(t)f_{\widehat\alpha}(\varepsilon^{-1}t)\chi(t)\times
f(\varepsilon^{-1/4}s),
\end{equation}
where $f_{\widehat\alpha}$ is the first eigenfunction of the
operator $H[a,m,\widehat\alpha;\widehat\eta]$ whose $L^2$-norm in
$\mathbb R$ is equal to $1$. The functions $\chi$ and $f$ are as in
(\ref{cu-off-Sec5}) and (\ref{f-Sec5}), and $a_0(t)$ is defined by~:
$$a_0(t)=1-t\kappa_0,\quad\text{ where }\kappa_0:=(\kappa_{\rm r})_{\rm max}.$$
Let us introduce the operator,
$$H_1=\partial_t+(t-\widehat\eta)^3-\xi^2(t-\widehat\eta).$$
Then, using the exponential decay of $f_{\widehat\alpha}$ and the
choice of gauge in Proposition~\ref{Agd1}, one gets (for the
detailed calculations, see \cite[Proposition~5.4.3]{kachTh}),
\begin{eqnarray}
&&\hskip-0.8cm\left|q_{\varepsilon,\mathbf F,\Omega}(u_\varepsilon)-
\varepsilon^{-2}\int_{\mathbb
R_+}\left(|f_{\widehat\alpha}'(t)|^2+|(t-\widehat\eta)f_{\widehat\alpha}(t)|^2
+\kappa_0\varepsilon(H_1f_{\widehat\alpha})(t)\,f_{\widehat\alpha}(t)\right)dt\right|\leq
C\varepsilon^{-1/2},\label{II-EHeMo10.25}\\
&&\hskip-0.8cm\left| q_{\varepsilon,\mathbf
F,\overline{\Omega}^c}(u_\varepsilon)-\varepsilon^{-2}\int_{\mathbb
R_-}\left(|f_{\widehat\alpha}'(t)|^2+|(t-\widehat\eta)f_{\widehat\alpha}(t)|^2
+\kappa_0\varepsilon(H_1f_{\widehat\alpha})(t)\,f_{\widehat\alpha}(t)\right)dt\right|\leq
C\varepsilon^{-1/2}.\label{II-EHeMo10.25'}
\end{eqnarray}
An integration by parts yields, thanks to (\ref{II-M1}) and
(\ref{II-C1}),
$$\int_{\mathbb
R_+}H_1\,f_{\widehat\alpha}(t)\,f_{\widehat\alpha}(t)\,dt+\frac1m
\int_{\mathbb
R_-}H_1\,f_{\widehat\alpha}(t)\,f_{\widehat\alpha}(t)\,dt=\widehat{\mathcal
C}_1\left(a,m\right).$$ Therefore, (\ref{II-EHeMo10.25}) and
(\ref{II-EHeMo10.25'}) together with our choice of
$f_{\widehat\alpha}$ yield the estimate,
$$
\left|\mathcal Q[\kappa,H](u_\varepsilon)
-\varepsilon^{-2}\left\{\mu_1\left(a,m,\widehat\alpha;\widehat\eta\right)
+\widehat{\mathcal
C}_1\left(a,m\right)\kappa_0\varepsilon\right\}\right|\leq
C\varepsilon^{-1/2}.$$ The decay of $f_{\widehat\alpha}$ at infinity
also yields,
$$\left|\|u_\varepsilon\|_{L^2(\mathbb R)}^2-1\right|\leq
C\exp\left(-\varepsilon^{-1}\right),\quad \forall
\varepsilon\in]0,\varepsilon_0].$$ The application of the min-max
principle now achieves the proof of the proposition.\hfill$\Box$\\

\begin{rem}\label{UB-cr-R}
Using Theorem~\ref{II-mu-th'-1}, we get a better version of the
upper bound~(\ref{UB-cr-eq}),
\begin{eqnarray}\label{II-eq-BS'}
\varepsilon^2\mu^{(1)}(\kappa, H)&\leq&
b_1\left(a,m\right)(\widehat\alpha-\alpha) +
\widetilde{\mathcal C}_1\left(a,m\right)(\kappa_{\rm r})_{\rm max}\,\varepsilon\\
&&+C\left(\varepsilon^{3/2}+|\widehat\alpha-\alpha|^2\right),\quad\forall
\varepsilon\in]0,\varepsilon_0],\nonumber
\end{eqnarray}
where $b_1(\cdot,\cdot)$ and $\widetilde{\mathcal C}_1(\cdot,\cdot)$
are defined in (\ref{II-C1a,m}) and (\ref{II-b0a,m}) respectively.
\end{rem}

In the next proposition, using the existence of ground states
(cf.~Proposition~\ref{ext-EF}),  we shall  determine a lower bound
for $\mu^{(1)}(\kappa,H)$.

\begin{prop}\label{LB-cr}
Under the above hypotheses and notations, there exist constants
$C,\varepsilon_0>0$, such that, if
$\varepsilon\in]0,\varepsilon_0]$, then,
\begin{equation}\label{LB-cr-eq1}
\varepsilon^2\mu^{(1)}(\kappa,H)\geq
b_1\left(a,m\right)(\widehat\alpha-\alpha) +\widetilde{\mathcal
C}_1\left(a,m\right)(\kappa_{\rm r})_{\rm
max}\,\varepsilon-C(\varepsilon^{4/3}+|\widehat\alpha-\alpha|^2).
\end{equation}
\end{prop}
\paragraph{\bf Proof.}
Let us  consider a  partition of unity
$\left(\chi_{j,\varepsilon^{1/3}}\right)_{j\in\mathbb Z^2}$ of
$\mathbb R^2$ that satisfies~:
\begin{equation}\label{sum}
\sum _{j\in J}|\chi _{j,\varepsilon^{1/3}}(z)|^2=1,\quad \sum _{j\in
J}|\nabla \chi _{j,\varepsilon^{1/3}}(z)|^2\leq C\varepsilon^{-2/3},
\end{equation}
\begin{equation}\label{support1}
\text{supp }\chi _{j,\varepsilon^{1/3}}\subset
j\varepsilon^{1/3}+[-\varepsilon^{1/3},\varepsilon^{1/3}]^2.
\end{equation}
We define the following set of indices~:
$$
J^1_{\tau(\varepsilon)}:=\{j\in \mathbb Z^2;\quad {\rm supp}\chi
_{j,\varepsilon^{1/3}}\cap \Omega\not=\emptyset,\quad {\rm
dist}(\text{supp } \chi _{j,\varepsilon^{1/3}},\partial\Omega)\leq
\tau(\varepsilon) \},
$$
where the number $\tau(\varepsilon)$ is defined by~:
\begin{equation}\label{eq.5.26}
\tau(\varepsilon)=\varepsilon^{2\delta},\quad \text{with}\quad \frac
16\leq\delta\leq\frac 12,
\end{equation}
and the number $\delta$ will be chosen suitably.\\
We consider also another scaled partition of unity in $\mathbb R$~:
\begin{equation}\label{CHeMo9.22}
\psi_{0,\tau(\varepsilon)}^2(t)+\psi_{1,\tau(\varepsilon)}^2(t)=1,\quad|\psi'_{j,\tau(\varepsilon)}(t)|\leq\frac{C}{\tau(\varepsilon)}
,\quad j=0,1,\end{equation}
\begin{equation}\label{CHeMo9.23}
{\rm
supp}\,\psi_{0,\tau(\varepsilon)}\subset\left[\frac{\tau(\varepsilon)}{20},+\infty\right[,\quad
{\rm
supp}\,\psi_{1,\tau(\varepsilon)}\subset\left]-\infty,\frac{\tau(\varepsilon)}{10}\right].
\end{equation}
Notice that, for each $j\in J^1_{\tau(\varepsilon)}$, the function
$\psi_{1,\tau(\varepsilon)}(t)\chi_{j,\varepsilon^{1/3}}(s,t)$ can
be interpreted, by means of boundary coordinates, as a function in
$\mathcal N_{t_0}$ (cf. (\ref{N-t0})). Moreover, each
$\psi_{1,\tau(\varepsilon)}(t)\chi_{j,\varepsilon^{1/3}}(s,t)$ is
supported in a rectangle
$$K(j,\varepsilon)=]-\varepsilon^{1/3}+s_j,s_j+\varepsilon^{1/3}[\times
[0,\varepsilon^{2\delta}[$$ near $\partial\Omega$. The role of
$\delta$ is then to control  the width of each rectangle
$K(j,\varepsilon)$.\\
Let $\phi_\kappa$ be an $L^2$-normalized ground state of $\mathcal
P[\kappa,H]$ whose existence was shown in Proposition~\ref{ext-EF}. Since
$\phi_\kappa$ decays exponentially away from the boundary, we get,
\begin{equation}\label{KaV.16}
\left|\sum_{j\in J_{\tau(\varepsilon)}^1}\mathcal
Q[\kappa,H](\psi_{1,\tau(\varepsilon)}\chi_{j,\varepsilon^{1/3}}\phi_\kappa)-
\varepsilon^{-4}\mu^{(1)}(\kappa, H)\right|\leq C\varepsilon^{-2/3}.
\end{equation}
The proof of (\ref{KaV.16}) follows actually that of \cite[Formulas
(10.4), (10.5), (10,6)]{HeMo3}, see \cite[Proposition~5.5.1]{kachTh}.\\
For each $j\in J^1_{\tau(\varepsilon)}$, we define a unique point
$z_j\in\partial\Omega$ by the relation $s(z_j)=s_j$. We denote then
by $\kappa_j=\kappa_{\rm r}(z_j)$, $a_j(t)=1-\kappa_jt$, and
$$
A^j(t)= -t\left(1-\frac{t}2\kappa_j\right),
$$
Let us consider the $k$-family of differential operators,
$$H_{\varepsilon,j,k}=-\varepsilon^4a_j^{-1}
\partial_t^2(a_j\partial_t)+(1+2\kappa_jt)(\varepsilon^2k-A^j)^2.$$
We denote by $H_{\varepsilon,j,k}^{a,m}$ the self-adjoint operator
on the space
$$L^2(]-\varepsilon^{2\delta},\varepsilon^{2\delta}[;a_j(t)dt)$$
defined by~:
$$
H_{\varepsilon,j,k}^{a, m,\widehat\alpha}=\left\{\begin{array}{l}
H_{\varepsilon,j,k}-\widehat\alpha,\quad t>0,\\
\frac1{m}H_{\varepsilon,j,k}+a\widehat\alpha,\quad t<0,
\end{array}\right.
$$
with domain~:
\begin{eqnarray*}
D(H_{\varepsilon,j,k}^{a,m,\widehat\alpha})&=&\{u\in
H_0^1(]-\varepsilon^{2\delta},\varepsilon^{2\delta}[;\quad
u_{|_{]-\varepsilon^{2\delta},0[}}\in
H^2(]-\varepsilon^{2\delta},0[),\\
&&\hskip0.2cmu_{|_{]0,\varepsilon^{2\delta}[}}\in
H^2(]0,\varepsilon^{2\delta}[),\quad u'(0_+)=\frac1{m}u'(0_-)\}.
\end{eqnarray*}
We introduce~:
\begin{equation}\label{eq.5.32}
\mu_1(H_{\varepsilon,j,k}^{a,m,\widehat\alpha}):= \inf _{k\in
\mathbb R} \inf {\rm Sp} (H_{\varepsilon,j,k}^{a,m,\widehat\alpha}).
\end{equation}
Then one gets from (\ref{KaV.16}) the following result,
\begin{equation}\label{eq.5.33}
\varepsilon^2\mu^{(1)}(\kappa,H) \geq \left(\inf _{j\in
J^1_{\tau(\varepsilon)}} \mu
_1(H_{\varepsilon,j,k}^{a,m,\widehat\alpha})\right)+\mathcal
O(\varepsilon^{4/3}).
\end{equation}
Let us explain briefly how we get (\ref{eq.5.33}) (the details of
the calculations are given in~\cite[Section~11]{HeMo3}). We express
each term $\mathcal Q
[\kappa,H](\psi_{1,\tau(\varepsilon)}\chi_{j,\varepsilon^{1/3}}\phi_\kappa)$
in boundary coordinates. We work with the local choice of gauge
given in Proposition~\ref{Agd1}. We expand now all the terms by
Taylor's Formula near $(s_j,0)$. After controlling the remainder
terms, thanks to the exponential decay of the ground states away
from the boundary, we apply a partial Fourier transformation in the
tangential variable
$s$ and we get finally the desired result (see \cite[Proposition~5.5.3]{kachTh}).\\
Putting,
$$\beta=\kappa_j,\quad\xi=-\varepsilon k,\quad h=\varepsilon^2,$$
and applying the scaling $\widetilde t=h^{-1/2}t$, one gets,
$$\mu_1(H_{\varepsilon,j,k}^{a,m,\widehat\alpha})=h\mu_1(\mathcal H_{h,\beta,\xi}^{a,m,\widehat\alpha}).$$ By applying
Theorem~\ref{II-tp-mop-M}\footnote{The optimal choice of $\delta$
which gives a remainder in accordance with (\ref{eq.5.33}) is $\delta=5/12$.\\
We have to notice also that $\widetilde C_1(a,m)$ is negative for
$m\geq m_0$.}, we finish
the proof of the theorem.\hfill$\Box$\\

\paragraph{\bf Proof of Theorem~\ref{k2Th3}.}
It results from (\ref{II-eq-BS'}), (\ref{LB-cr-eq1}) and the
definition of $H_*(\kappa)$,
$$b_1\left(a,m\right)(\widehat\alpha-\alpha)
+\widetilde{\mathcal C}_1\left(a,m\right)(\kappa_{\rm r})_{\rm
max}\varepsilon+\mathcal
O\left(\varepsilon^{4/3}+|\widehat\alpha-\alpha|^2\right)=0.$$ This
yields,
$$|\widehat\alpha-\alpha|\leq C\varepsilon,$$
and consequently,
$$H_*(\kappa)=\frac{\kappa}\alpha-\frac{\widetilde{\mathcal
C_1}\left(a,m\right)}{b_1\left(a,m\right)\alpha^{3/2}}(\kappa_{\rm
r})_{\rm max}+\mathcal O(\kappa^{-1/3}).$$ This  achieves the proof
of the theorem upon setting,
\begin{equation}\label{C1-Fsec}\mathcal
C_1(a,\cdot)=-\frac{\widetilde{\mathcal
C}_1(a,\cdot)}{b_1(a,\cdot)}.\end{equation}

\section*{Acknowledgements}
I am deeply grateful to Professor B.~Helffer for his help and kind
encouragement during the preparation of this work. This work is
partially supported by the European Research Network `Postdoctoral
Training Program in Mathematical Analysis of Large Quantum Systems'
with contract number HPRN-CT-2002-00277, by the ESF Scientific
Programme in Spectral Theory and Partial Differential Equations
(SPECT) and by the `{\it Agence universitaire de la francophonie'}
(AUF).
\appendix
\section{Proof of Theorem~\ref{k2Re1}}\label{App-A}
We bring in this appendix the proof of the asymptotics
(\ref{k2Re1-eq3}). Notice that the Euler-Lagrange equations
associated to the functional (\ref{k2Re1-eq2}) has a solution
$(0,\mathbf A)$ that satisfies,
$${\rm curl}\,\mathbf A=1,\quad{\rm div}\,\mathbf A=0\text{ in
}\Omega,\quad \nu\cdot\mathbf A=0\text{ on }\partial\Omega,$$ and
the normal state $(0,\mathbf A)$ is unique. Define the following eigenvalue,
\begin{equation}\label{EV-Kach1}
\lambda^{(1)}(\delta,\gamma_0;\varepsilon)=\inf_{u\in
H^1(\Omega),u\not=0}
\frac{q_{\varepsilon}^{\delta,\gamma_0}(u)}{\|u\|_{L^2(\Omega)}^2},
\end{equation}
where,
$$q_{\varepsilon}^{\delta,\gamma_0}(u)=
\|(\nabla-i\varepsilon^{-2}\mathbf
A)u\|^2_{L^2(\Omega)}+\kappa^\delta\gamma_0\|u\|_{L^2(\partial\Omega)}^2,$$
and $\varepsilon=\frac1{\sqrt{\kappa H}}$. We define the following
upper critical field,
$$H_{C_3}(\delta,\gamma_0;\kappa)=\inf\{H>0;\quad
\lambda^{(1)}(\delta,\gamma_0;\varepsilon)\geq \kappa^2\}.$$ As for
(\ref{HC3=H*}), we can also show that the critical field,
\begin{equation}\label{HC3=H*-Ap}
H_*(\delta,\gamma_0;\kappa)=\inf\{H>0;\quad
\lambda^{(1)}(\delta,\gamma_0;\varepsilon)=\kappa^2\},
\end{equation}
is equal to the upper critical field for large values of $\kappa$.
Following the generalization of the analysis of
Helffer-Morame~\cite{HeMo3} in \cite{Kach1}, we obtain the following
asymptotics for $\lambda^{(1)}(\delta,\gamma_0;\varepsilon)$.

\begin{prop}\label{Ap-Kach1}
Let $H=H_*(\kappa,\delta,\gamma_0)$. The following asymptotics holds
as $\varepsilon$ tends $0$,
\begin{equation}\label{Ap-Kach1-eq1}
\varepsilon^2\lambda^{(1)}(\delta,\gamma_0;\varepsilon)=
\Theta\left(\varepsilon^2\kappa^\delta\gamma_0\right)\left(1+o(1)\right),
\end{equation}
where $\Theta(\cdot)$ is defined in (\ref{Th-gam}).
\end{prop}
\paragraph{\bf Proof.} We split the proof in two steps corresponding to the
determination of an upper bound and of a lower
bound.\\
{\it Step 1. Upper bound.}\\
We establish the following upper bound,
\begin{equation}\label{UB-App}
\varepsilon^2\lambda^{(1)}(\delta,\gamma_0;\varepsilon)\leq
\Theta\left(\varepsilon^2\kappa^\delta\gamma_0\right)\left(1+o(1)\right)\quad
(\varepsilon\to0).
\end{equation}
We have two cases to deal with,
\begin{center}
either
$\displaystyle\lim_{\varepsilon\to0}\varepsilon^2\kappa^\delta=+\infty$,
or
$\displaystyle\lim_{\varepsilon\to0}\varepsilon^2\kappa^\delta<+\infty$.
\end{center}
{\it Case 1.
$\displaystyle\lim_{\varepsilon\to0}\varepsilon^2\kappa^\delta=+\infty$
and $\gamma_0>0$.} In this case, by the min-max principle and
(\ref{HeMo3-ND-eq1}), we get,
$$\varepsilon^2\lambda^{(1)}(\delta,\gamma_0;\varepsilon)\leq
\varepsilon^2\mu^D(\varepsilon;\Omega)\leq 1+C\varepsilon^3.$$ We
obtain then the upper bound (\ref{UB-App}) upon recalling that
$\Theta(\gamma)$ is exponentially close to $1$ when
$\gamma\to+\infty$ (cf. \cite{Kach1}).\\
{\it Case 2.
$\displaystyle\lim_{\varepsilon\to0}\varepsilon^2\kappa^2<+\infty$
or $\gamma_0\leq0$.} Let $\eta:=\varepsilon^2\kappa^\delta\gamma_0$.
We construct the following quasi-mode by means of boundary
coordinates (cf. Subsection~\ref{sbS-BC}),
$$u_\varepsilon(s,t)=\varepsilon^{-3/4}a^{-1/2}\exp\left(-i\frac{\xi(\eta)
s}{\varepsilon}\right) \varphi_\eta\left(\varepsilon^{-1}t\right)
\chi(t)\times f\left(\varepsilon^{-1/2}s\right),$$ where the
functions $\chi$ and $f$ are as in (\ref{cu-off-Sec5}) and
(\ref{f-Sec5}) respectively. Then following \cite[Proof of
Proposition~3.1]{Kach1}, we get the upper bound (\ref{UB-App}).\\
{\it Step 2. Lower bound.}\\
We establish the following lower bound,
\begin{equation}\label{LB-App}
\varepsilon^2\lambda^{(1)}(\delta,\gamma_0;\varepsilon)\geq
\Theta\left((1-C\varepsilon^{1/2})\varepsilon^2\kappa^\delta\right)-C\varepsilon^{5/2}.
\end{equation}
We have, thanks to (\ref{IMS-f}), (\ref{P-O-U-eq2}),
(\ref{HeMo3-int}) and (\ref{HeMo3-bnd}) (with the choice $\rho=3/4$
and $\theta=1/4$),
$$
q_{\varepsilon}^{\delta,\gamma_0}(u)\geq\varepsilon^{-2}
\left(\sum_{\rm int}\|\chi_j^\varepsilon u\|^2
+\left(1-C\varepsilon^{1/2}\right)\sum_{\rm
bnd}q^{\delta,\widetilde\eta}_{\varepsilon,\mathbb R\times\mathbb
R_+}(\chi_j^\varepsilon u)-C\varepsilon^{1/2}\|u\|^2\right),
$$
where $\widetilde\eta=(1-C\varepsilon)^{-1}\eta$. Applying the
min-max principle, we get the lower bound (\ref{LB-App}). We achieve
now the proof of the theorem upon recalling the asymptotic behavior
of
$\Theta(\cdot)$ (cf.~\cite{Kach1}).\hfill$\Box$\\

\begin{lem}\label{app-pr}
There exists a unique $\eta_0<0$ such that $\Theta(\eta_0)=0$.
Moreover, given $\gamma_0\in\mathbb R$, there exists a unique
$\ell(\gamma_0)>0$ such that
$\Theta(\gamma_0\cdot\ell(\gamma_0))=\ell(\gamma_0)^2$.
\end{lem}
\paragraph{\bf Proof.}
The existence and uniqueness of $\eta_0$ comes from the monotonicity
of $\Theta(\cdot)$ and the behavior of
$\Theta(\cdot)$ at $-\infty$ ($\Theta(\gamma)\sim-\gamma^2$).\\
Let us define the function $h(\eta)=\Theta(\gamma_0\eta)-\eta^2$.
Notice that $h(0)=\Theta_0>0$ and $h(\eta)<0$ for all $\eta\geq1$.
Therefore, thanks to the Intermediate Value Theorem, there exists a
solution $\ell(\gamma_0)$ of $h(\ell(\gamma_0))=0$ and this solution
is in $]0,1[$. Using (\ref{Th'-gam}), we get,
$$h'(\eta)=\gamma_0|\varphi_{\gamma_0\eta}(0)|^2-2\eta,$$
where $h'(\eta)<0$ in $]0,1[$ if $\gamma_0\leq0$. If $\gamma_0>0$,
then thanks to the Min-Max Principle, (\ref{xi-gam}) and
(\ref{l1-ga,xi}),
$$\gamma_0\eta|\varphi_{\gamma_0\eta}(0)|^2\leq
\Theta(\gamma_0\eta)-\Theta_0,\quad \Theta_0<\ell(\gamma_0)<1,$$ and
consequently, (recall that $\Theta_0>\frac12$),
$$h'(\eta)\leq
\frac{\Theta(\gamma_0\eta)-1}{\eta}-2\eta\leq\frac{1-\Theta_0}\eta-2\eta<0\quad\text{in
}]\Theta_0,1[.$$
Therefore, $\ell(\gamma_0)$ is unique.\hfill$\Box$\\

\paragraph{\bf Proof of Theorem~\ref{k2Re1}.}
Using (\ref{HC3=H*-Ap}) and Proposition~\ref{Ap-Kach1}, we have to
analyze the limit of $\frac{\kappa^\delta}{(\kappa
H_*(\kappa))^{1/2}}$ as $\kappa\to+\infty$. Let
\begin{equation}\label{eq1-lim}
\beta=\lim_{\kappa\to+\infty}\frac{\kappa^\delta}{(\kappa
H_*(\kappa))^{1/2}}, \end{equation} with $0\leq\beta\leq+\infty$.
Notice that, thanks to (\ref{Ap-Kach1-eq1}) and to the definition of
$H_*(\kappa)$, we have the following equation,
\begin{equation}\label{eq-Bas}
\Theta\left(\frac{\kappa^\delta}{(\kappa
H_*(\kappa))^{1/2}}\gamma_0\right)=\frac{\kappa}{H_*(\kappa)}(1+o(1))\quad
\text{as }\kappa\to+\infty.
\end{equation}
The above equation gives, thanks to the monotonicity of
$\Theta(\cdot)$ and the definition of $\eta_0$,
\begin{equation}\label{eq1-App-pr}
\beta\gamma_0\geq \eta_0.
\end{equation}
{\it Case 1. $\delta<1$.}\\
We show in this case that $\beta=0$. Suppose by contradiction that
$\beta>0$. If $\beta<+\infty$, then, thanks to (\ref{eq1-lim}),
$$H_*(\kappa)=\frac1{\beta^2}\kappa^{2\delta-1}(1+o(1)).$$
Substituting in (\ref{eq-Bas}), we get,
$$\Theta(\beta\gamma_0)=\frac1{\beta^2}\kappa^{2(1-\delta)}(1+o(1)),$$
which is impossible since $\delta<1$. If $\beta=+\infty$, then,
thanks to (\ref{eq1-App-pr}), this case is possible only if
$\gamma_0>0$. Using (\ref{eq-Bas}), we get, thanks to the decay of
$\Theta(\cdot)$ at $+\infty$,
$$\frac{\kappa}{H_*(\kappa)}=1+o(1),$$
and consequently,
$$\frac{\kappa^\delta}{(\kappa
H_*(\kappa))^{1/2}}=\mathcal O(\kappa^{\delta-1}),$$ which is a
contradiction since $\delta<1$.\\
{\it Case 2. $\delta=1$.}\\
If $\beta=+\infty$, we get a contradiction as in the above case.
Therefore, $\beta<+\infty$. Then, combining (\ref{eq1-lim}) and
(\ref{eq-Bas}), we get $\Theta(\beta\gamma_0)=\beta^2$. Thus, by
Lemma~\ref{app-pr}, $\beta=\ell(\gamma_0)$.\\
{\it Case 3. $\delta>1$ and $\gamma_0>0$.}\\
It is sufficient to prove that $\beta=+\infty$. Suppose by
contradiction that $\beta<+\infty$. Then, (\ref{eq-Bas}) will give
$H_*(\kappa)=\kappa(1+o(1))$ while (\ref{eq1-lim}) will give
$$H_*(\kappa)=\frac1{\beta^2}\kappa^{2\delta-1}(1+o(1)),$$ which is
impossible since $\delta>1$.\\
{\it Case 4. $\delta>1$ and $\gamma_0<0$.}\\
In this case $\beta$ should be  finite, thanks to
(\ref{eq1-App-pr}). We deduce then from (\ref{eq1-lim}) that
$$H_*(\kappa)=\frac1{\beta^2}\kappa^{2\delta-1}(1+o(1)).$$ Since
$\delta>1$, (\ref{eq-Bas}) gives~: $\Theta(\beta\gamma_0)=0$.
Therefore, by Lemma~\ref{app-pr},
$\beta\gamma_0=\eta_0$.\hfill$\Box$

\end{document}